\documentclass[final,3p,times,review]{elsarticle} 
\pdfoutput=1
\usepackage{amssymb,amsmath,array} 

\usepackage{float} 

\usepackage[font=normal]{caption} 

\usepackage[flushleft]{threeparttable} 
\usepackage{multirow}
\usepackage{lscape}
\usepackage{pdflscape}
\usepackage{amsmath}
\usepackage{makecell}

\usepackage{subcaption}
\usepackage{graphicx}

\usepackage{nomencl}
\makenomenclature
\usepackage{etoolbox}
\renewcommand\nomgroup[1]{%
	\item[\bfseries
	\ifstrequal{#1}{A}{Symbols}{%
		\ifstrequal{#1}{B}{Subscripts}{%
			\ifstrequal{#1}{C}{Abbreviations}{}}}%
	]}

\usepackage{soul} 

\usepackage{color} 

\usepackage{xcolor} 

\definecolor{mc1}{rgb}{1,0,0} 

\journal{} %Energy \& Environmental Science

\usepackage{lineno}

\begin{document}

\begin{frontmatter}

%\title{Dynamic responses of solid oxide cells (SOC) to changes in voltage/current: Characteristic times of transport phenomena} 
\title{Characteristic time of transient response of solid oxide cells (SOCs) to changes in voltage/current: from theory to applications}
\author[add1]{Zhaojian Liang} 
\author[add2]{Jingyi Wang}
\author[add1]{Liang An} 
\author[add3]{Yang Wang}
\author[add3]{Meng Ni}
\author[add1]{Mengying Li\corref{cor1}} 
\ead{mengying.li@polyu.edu.hk} 
\cortext[cor1]{Corresponding author} 
%\fntext[fn1]{The two authors have the same contribution to this study.} 

\address[add1]{Department of Mechanical Engineering \& Research Institute for Smart Energy, The Hong Kong Polytechnic University, Hong Kong SAR}
\address[add2]{School of Science, Harbin Institute of Technology, Shenzhen, 518055, China}
\address[add3]{Department of Building and Real Estate, The Hong Kong Polytechnic University, Hong Kong SAR}

\begin{abstract} 
The intermittency of solar and wind power can be addressed by integrating them with Solid Oxide Cells (SOCs). This study delves into the transient characteristics of SOCs and their dependence on dynamic heat and mass transfer processes. Non-dimensional analysis was used to identify influential parameters, followed by a 3-D numerical simulation-based parametric analysis to examine the dynamic gaseous and thermal responses of SOCs with varying dimensions, material properties, and operating conditions. For the first time, we proposed characteristic times $\tau_{\rm h}$ and $\tau_{\rm m}$ to describe the relationship between SOC transients and multiple parameters. These characteristic times represent the overall heat and mass transfer rats in SOCs. Their effectiveness was validated against literature and demonstrated potential in characterizing the transient characteristics of other electrochemical cells. Besides, two examples are provided to illustrate how $\tau_{\rm h}$ and $\tau_{\rm m}$ facilitate SOC design and control at minimal computational cost. 
\end{abstract}

\begin{highlights} 
\item Investigates the relationship between SOC transients and multiple parameters.
\item 3-D numerical simulation examines gaseous and thermal responses of SOCs.
\item Characteristic times $\tau_{\rm h}$ and $\tau_{\rm m}$ generalize heat and mass transfer rates. 
\item Proposed characteristic times validated against literature. 
\item Characteristic times facilitate SOC design and control at minimal computational cost.

\end{highlights}


\end{frontmatter} 

\section{Introduction} 
Utilizing renewable energy sources is a crucial path to relief energy shortage and environmental crisis. According to the BP Energy Outlook 2023, solar and wind power capacity is expected to increase four-fold by 2030 from 2019 levels \cite{2023EnerOutlook}. However, the integration of solar and wind power is a burden for the electric grid due to their intermittent and unstable nature. To achieve power load balancing in the regime of high renewable penetration, it is necessary to develop large-scale energy storage and conversion devices such as SOC \cite{WANG2023SciBul}. SOCs can operate reversibly in both electrolysis and fuel cell modes \cite{myung_switching_2016}. In electrolysis mode, solid oxide electrolysis cells (SOECs) can store the intermittent solar \cite{SUN2023125725} and wind power \cite{WANG2019255} in alternative fuels such as hydrogen and syngas \cite{EVELOY2019550}. In fuel cell mode, solid oxide fuel cells (SOFCs) can generate power from various fuels \cite{XU2022VariousFuel,HUANG2022119018}. Furthermore, SOCs have superior electrical efficiency compared to existing electrochemical cells, regardless of whether they operate in SOEC or SOFC mode \cite{BUTTLER20182440,XU2022VariousFuel,EVELOY2019550}. Due to the high flexibility and efficiency, SOC technology has great potential in addressing the intermittency issues associated with renewable power generation.

Considering the application scenarios, SOCs need to operate under unsteady conditions. However, rapid load changes can result in electrical overshoots or undershoots due to the slow responses of heat and mass transfer in SOCs \cite{LIANG2023116759,LUO2015637}. On one hand, the transient current-voltage characteristics of SOEC may impair the operational efficiency of the integrated systems that combine SOECs with intermittent power sources \cite{LIANG2023116759}. On the other hand, the load changes due to varying demands may induce under-performance, thermal safety issues, and degradation \cite{WANG2022ThermalCycle,XU2022VariousFuel,ZHU2022136159,NERAT2017728} of SOFC. To address these issues, a number of studies \cite{LIU2022115318,ALBRECHT2016402,PREININGER2019113695,EICHHORNCOLOMBO2020117752,FOGEL20199188} have proposed control strategies for safe operation and optimized design parameters to improve the inherent transient characteristics of SOCs. For instance, Nerat~\cite{NERAT2017728} suggested that increasing the thickness of the anode supporting layer to 0.1\,mm can prevent local fuel starvation during dynamic operation of a planar SOFC, and slowing the load variation rate to a time constant of 50\,ms can reduce electrical overshoot. Bae et al.~\cite{BAE2019112152} emphasized the importance of fuel utilization on the current relaxation time of SOFC after electrical load changes.  These studies can provide valuable qualitative insights into SOC transients. However, their quantitative findings are rarely applicable to other SOCs due to the differences in specifications and operating conditions. The variety of SOC transients constrains the generalizability of SOC control and design. 

To establish a generalized relationship between the transient characteristics and parameters of SOCs, considerable efforts have been made to identify a characteristic time constant that governs the SOC transients. Some studies \cite{ACHENBACH1995105,MENON2012227,SERINCAN2009864,NERAT2017728} held the view that the electrical responses of SOCs depend on the time constants of heat and mass diffusion in the thickness direction of SOC, i.e., $\delta^2 / \alpha$ and $\delta^2 / D$. While a few studies \cite{BAE2018405,LIANG2023116759} argued that the advection time constant $L/V$ is also important for SOC transients. Unfortunately, neither of these two views can fully and quantitatively explain the dynamic responses of SOCs, which depend on multiple parameters such as cell length \cite{BANERJEE2018996}, electrode thickness \cite{NERAT2017728,BAE2019112152}, and inlet flow rates \cite{BAE2019112152,LIU2022115318}. In other words, existing knowledge is insufficient to describe the relationship between SOC transients and multiple parameters in a generalized way. This research gap results in significant challenges in the design and control of SOCs, which typically involve tens of adjustable parameters. This hinders the development of SOC technology for renewable energy storage and conversion. In this study, we address this critical research gap by investigating the characteristic time of dynamic response of SOCs from both theoretical and practical perspectives. Our work provides new insights into the relationship between SOC transients and multiple parameters, paving the way for more efficient design and control of SOCs. This has significant implications for the integration of SOC technology with renewable energy sources, with the potential to accelerate the transition to a sustainable energy future.

In Section\,\ref{Sec:metho}, non-dimensional analysis is applied to screen the parameters that influence the heat and mass transfer in SOC, and then a parametric study is conducted by 3-D transient simulation. In Section\,\ref{Sec:parametric}, based on the parametric study, generalized characteristic times for heat and mass transfer in SOC are firstly proposed to precisely describe the relationship between the step-response time and SOC parameters (i.e., dimensions, material properties, and operating conditions) with clear physics meanings. The effectiveness the proposed characteristic times is also validated against the transient characteristics of multiple types of SOCs and a PEMFC reported in the literature. Finally, in Section\,\ref{Sec:application}, two examples are provided to illustrate the applications of the proposed characteristic times to the design and control of SOC transients. Furthermore, the conclusions drawn from this study may be extended to other electrochemical cells.

\section{Methodology} \label{Sec:metho}
\begin{figure}[h]
    \includegraphics[width=1 \textwidth]{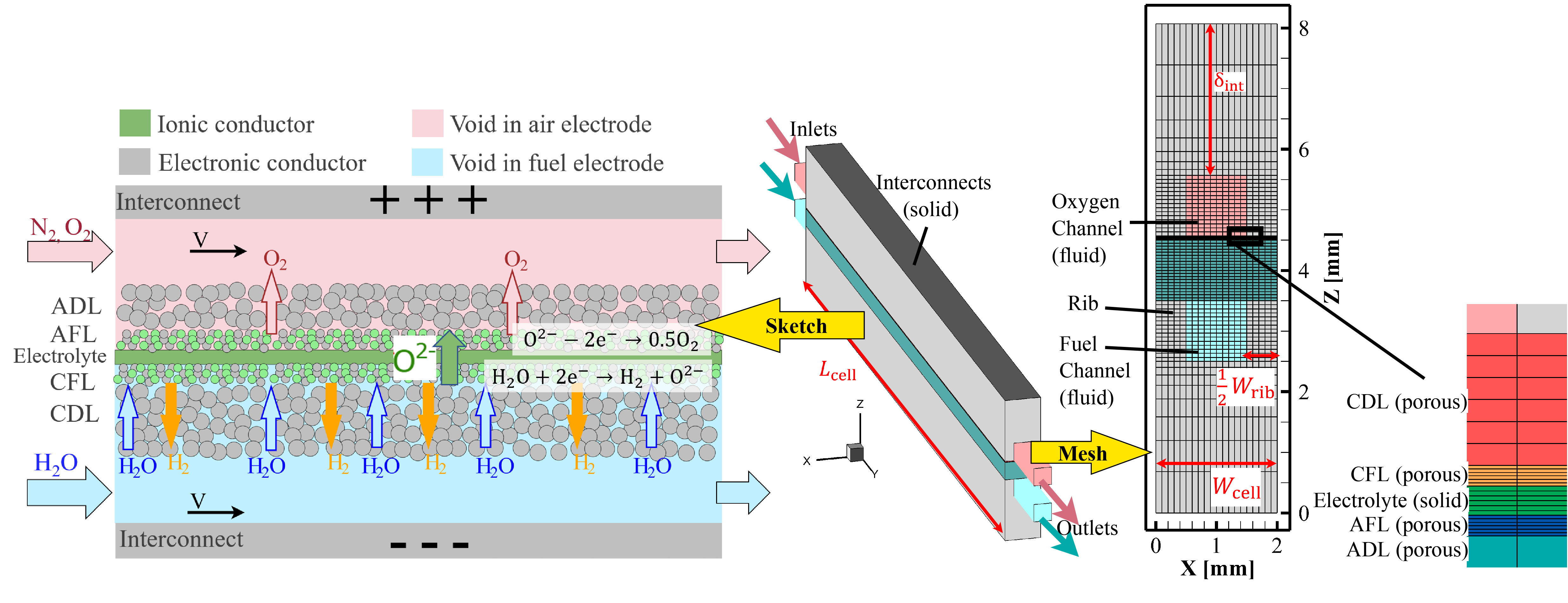} 
    \caption{Geometry and working principle of SOEC. }
    \label{Fig:sketch-geo} 
\end{figure}

In SOEC mode, electrochemical reactions consume both steam and electrical power. Conversely, in SOFC mode, hydrogen and oxygen are consumed while water and electrical power are generated. Figure~\ref{Fig:sketch-geo} illustrates the working principle of SOEC. The top and bottom interconnects serve as electron conductors. The porous media, including the Anode Diffusion Layer (ADL), Anode Functional Layer (AFL), Cathode Functional Layer (CFL), and Cathode Diffusion Layer (CDL), conduct electrons and contain voids that allow for gas transport. Oxygen ion transport occurs exclusively in the solid oxide electrolyte, AFL, and CFL at high temperatures. In SOEC mode, hot steam from the fuel channel diffuses through the CDL to the CFL where electrochemical reactions occur at triple-phase boundaries, splitting steam into hydrogen and oxygen ions. Oxygen ions then transport across the solid oxide electrolyte to the AFL where oxygen is generated through electrochemical reactions. The transient characteristics of SOC are heavily influenced by the transport processes of electrons, ions, gases, and heat.

\subsection*{Numerical modeling}
This study conducts a comprehensive parametric analysis through numerical simulation to identify the key parameters that influence the transient characteristics of SOEC. The three-dimensional Computational Fluid Dynamics (CFD) model utilized for simulation has been previously validated \cite{LIANG2023116759}. Fig.\,\ref{Fig:sketch-geo} demonstrates the geometry and mesh of the cathode-supporting SOEC used in this study. The adoption of a 3-D single-channel model reduces computational cost while maintaining high spatial resolution \cite{BAE2019112152,Su2022}. Fick’s law and the Butler-Volmer equation are employed to model gas diffusivity and electrochemical reactions, respectively. Conservation equations for momentum, species mass, energy, electronic charge, and ionic charge are coupled and solved to obtain transient field data for velocity, species mass fractions, temperature, and electrical and ionic potentials within the SOEC. An adaptive time-stepping algorithm \cite{LIANG2023116759} is applied to calculate the time-stepping size of the transient simulation based on variations in species concentration and temperature. This ensures sufficient temporal resolution to resolve electrical behaviors induced by heat and mass transfer. Further details of the numerical model can be found in Ref.\,\cite{LIANG2023116759}.

\subsection*{Non-dimensional analysis}
Non-dimensional analysis \cite{BookIncompressible} is a widely used technique in Fluid Mechanics that allows for the generalization of important information about similar physical phenomena. A few studies \cite{Dim_WEN20117519,Dim_NIELSEN2021229675,Dim_SAYADIAN2021228997,Dim_SAYADIAN2022122557} have employed non-dimensional analysis to study 1-D and 2-D models of SOC and have gained a deeper understanding of steady-state transport phenomena within SOCs. Achenbach \cite{ACHENBACH1995105} investigated the relationship between the electrical responses of SOFC and dimensionless time scale of heat transfer. In this study, we applied non-dimensional analysis to a 3-D transient SOC model for the first time to investigate transient heat and mass transfer. By analyzing the conservation equations of energy and species through non-dimensional analysis, we identified the variables and time constants that directly influence SOC transients. 

To perform non-dimensional analysis on the heat and mass transfer in SOC, assumptions are made to simplify the governing equations used in the numerical models. Fluid properties are assumed to remain constant in non-dimensional analysis, allowing properties such as density $\rho$, diffusivity $D$, thermal conductivity $k$, and specific heat $c_p$ can be taken out of the partial differential signs. Also, the thermal transport term due to mass diffusion is  neglected in energy equations for simplicity. Table\,\ref{Tab:Dimensional} presents the simplified governing equations, variable scales, and  non-dimensional equations. We introduced two characteristic times $\tau_{\rm m}$ and $\tau_{\rm h}$ to scale the response times of mass and heat, respectively. The mathematical expressions for $\tau_{\rm m}$ and $\tau_{\rm h}$ will be determined and discussed in Section\,\ref{Sec:parametric}. Other variables are non-dimensionalized as follow:
\begin{itemize}
    \item The variation of the mass fraction $Y_i$, where $i$ denotes the species of interest, is scaled by the difference between its initial mass fraction at $t=0$ and its final steady state mass fraction at $T_{t\rightarrow \infty}$.  This scaling method is also applied to obtain the dimensionless mole fraction $X_{i}^*$, and dimensionless temperature $T^*$. 
    \item Spatial dimensions are scaled by the corresponding dimensions of the cell: $y$ by the length of the cell $L_{\rm cell}$, $x$ by the width of the cell $W_{\rm cell}$, $z$ by the thickness or height of the domain. 
    \item The inlets of anode and cathode fluid channels are selected as the reference points. Velocity $V$ is scaled by the inlet velocity $V_{\rm in}$. Fluid properties such as $D$, $\rho$, $c_p$, and $k$ are scaled by their respective reference properties under  inlet pressure, temperature, and species composition, denoted with a subscript `0'. Similarly, solid component properties (subscript `s') and porous media properties (subscript `eff') can be scaled in the same manner.
\end{itemize}

According to the dimensionless equations presented in Table,\ref{Tab:Dimensional}, the variations of species mass fraction $Y$ and temperature $T$, which reflect the mass and heat transfer in SOC, are influenced by: (1) cell dimensions,i.e., $W_{\rm cell}$, $L_{\rm cell}$, $H_{\rm ch}$, $\delta_{\rm ADL}$, $\delta_{\rm AFL}$, $\delta_{\rm CDL}$, and $\delta_{\rm CFL}$; (2) operating conditions, i.e., flow conditions $V_{\rm in}$, $T_{\rm in}$, and $X_{\rm H_2O/O_2/H_2, in}$, and electrical conditions denoted as current $i$ and voltage $U$ that determine the heat and mass sources; (3) fluid properties (i.e., $D$, $\rho$, $c_p$, and $\alpha$), and material properties (i.e., $\rho_{\rm s}$, $c_{p,\rm s}$, $k_{\rm s}$, $\alpha_{\rm s}$, $\varepsilon_{\rm ADL}$, $\varepsilon_{\rm AFL}$, $\varepsilon_{\rm CDL}$, and $\varepsilon_{\rm CFL}$). Note that, fluid properties and operating conditions are interdependent since fluid properties are influenced by operating conditions such as $T_{\rm in}$ and $X_{\rm H_2O/O_2/H_2, in}$. 

Besides, in Table\,\ref{Tab:Dimensional}, we identified the time constant ratios (bold terms) that govern the solutions of $Y_i^*$ and $T^*$. Each time constant ratio has a denominator representing the time scale of a local transport phenomenon in a specific location. For example, $\delta_{\rm{DL}}^2 / D_{\rm{eff,0}}$ represents the gas diffusion time constant across the thickness of DL, $H_{\rm{ch}}^2 / \alpha_0$ represents the thermal diffusion time constant across the height of channel, and $L_{\rm{cell}} / V_{\rm in}$ is the advection time constant. In addition, $\tau_{\rm m}$ or $\tau_{\rm h}$ serves as the numerator of all time constant ratios, representing the global time scale of mass/heat transfer in the entire SOC. Apparently, the expressions of $\tau_{\rm m}$ and $\tau_{\rm h}$ are critical for generalizing SOC transient characteristics but remain unknown in non-dimensional analysis. In Section\,\ref{Sec:parametric}, we will perform a parametric study on the SOC transients by varying the dimension parameters, operating conditions, and material properties in simulation. Based on the results, we will derive the expressions for the characteristic times of heat and mass transfer $\tau_{\rm h}$ and $\tau_{\rm m}$ and validate our proposed characteristic time with literature. In Section\,\ref{Sec:application}, we will discuss applications of characteristic times. 

\begin{table}[]
\centering
\small
\caption{Non-dimensional analysis on the heat and mass transfer in SOC.}
\label{Tab:Dimensional}
\begin{threeparttable}
\begin{tabular}{ll}
\hline
Simplified governing equations for non-dimensional analysis                       & Domains         \\ \hline
\textit {Conservation of species} &                \\
$\left(\frac{\partial Y_i}{\partial t}+\vec{V} \cdot \nabla Y_i\right)- D \nabla^2 Y_i=0 $                            & Fluid channel \\
$\left(\frac{\partial\left(\varepsilon Y_i\right)}{\partial t}+\vec{V} \cdot \nabla Y_i\right)-  D_{\rm{eff}} \nabla^2 Y_i= 0$                            & ADL,CDL        \\
$\left(\frac{\partial\left(\varepsilon Y_i\right)}{\partial t}+\vec{V} \cdot \nabla Y_i\right)-  D_{\rm{eff}} \nabla^2 Y_i= \frac{S_{{\rm m},i}}{\rho} $                            & AFL,CFL        \\
\textit {Conservation of energy}  &                \\
$\frac{\partial T}{\partial t}+ \vec{V} \cdot \nabla T-\nabla \cdot(\alpha \nabla T)=0$                            & Fluid channel \\
$\frac{\partial T}{\partial t}+ \frac{\rho c_p}{\rho_{\rm{eff}} c_{p,\rm{eff}}} \vec{V} \cdot \nabla T-\nabla \cdot\left(\alpha_{\rm{eff}} \nabla T\right)=\frac{S_{\rm{h}}}{\rho_{\rm{eff}} c_{p,\rm{eff}}}$                            & Porous\tnote{1}         \\
$\frac{\partial T}{\partial t}-\nabla \cdot\left(\alpha_{\rm{s}} \nabla T\right)=\frac{S_{\rm{h}}}{\rho_{\rm{s}} c_{p,\rm s}}$                            & Solid \tnote{2}   \\ \hline
Scaling \& definitions\tnote{3}                          &                \\
\multicolumn{2}{l}{\makecell[l]{$t^*=\frac{t}{\tau_{\rm m}}$ or $\frac{t}{\tau_{\rm h}}$, $Y_i^*=\frac{Y_i-Y_{i,t=0}}{\Delta Y_i}=\frac{Y_i-Y_{i,t=0}}{Y_{i,t\rightarrow \infty}-Y_{i,t=0}}$, $X_i^*=\frac{X_i-X_{i,t=0}}{X_{i,t\rightarrow \infty}-X_{i,t=0}}$, $V^*=\frac{V}{V_{\rm in}}$, $D^*=\frac{D}{D_0}$, $D_{\rm eff}^*=\frac{D_{\rm eff}}{D_{\rm eff,0}}$, $T^* =\frac{T-T_{t=0}}{\Delta T}=\frac{T-T_{t=0}}{T_{t\rightarrow \infty}-T_{t=0}}$, 
\\
$\nabla^* = L_{\rm cell}\nabla$, $x^*=\frac{x}{W_{\rm cell}}$, $y^*=\frac{y}{L_{\rm cell}}$, $z^*=\frac{z}{H_{\rm ch}}$ or $\frac{z}{\delta_{\rm DL}}$ or $\frac{z}{\delta_{\rm FL}}$, $S_{{\rm m},i}^* = \frac{S_{{\rm m},i} \tau_{\rm m}}{\rho_0 \Delta Y_i}$, $S_{\rm h}^* = \frac{S_{\rm{h}} \tau_{\rm h}}{\rho_0 c_{p, 0} \Delta T}$ or $\frac{S_{\rm{h}} \tau_{\rm h}}{\rho_{\rm s} c_{p,\rm s} \Delta T}$
\\
$\rho^*=\frac{\rho}{\rho_{\rm 0}}$, $\rho_{\rm eff}^*=\frac{\rho_{\rm eff}}{\rho_{0}}$, $c_p^* = \frac{c_p}{c_{p,0}}$, $c_{p,\rm eff}^* = \frac{c_{p,\rm eff}}{c_{p,0}}$, $k^* = \frac{k}{k_0}$, $k_{\rm eff}^* = \frac{k_{\rm eff}}{k_{\rm eff,0}}$, $k_{\rm s}^* = \frac{k_{\rm s}}{k_{\rm s,0}}$, $\alpha_0=\frac{k_0}{\rho_0 c_{p,0}}$, $\alpha_{\rm eff,0}=\frac{k_{\rm eff,0}}{\rho_{\rm eff,0} c_{p,{\rm eff,0}}}$, $\alpha_{\rm s,0}=\frac{k_{\rm s}}{\rho_{\rm s} c_{p,{\rm s}}}$  }}                      \\ \hline
Dimensionless equations           & Domains         \\ \hline
\textit {Conservation of species} &                \\
$\frac{\partial Y_i^*}{\partial t^*}+\boldsymbol{\frac{\tau_{\rm{m}}}{L_{\rm{cell}} / V_{\rm in}}} \vec{V}^* \cdot \nabla^* Y_i^*-D^*\left(\boldsymbol{\frac{\tau_{\rm m}}{W_{\rm{cell}}^2 / D_0}} \frac{\partial^2 Y_i^*}{\partial x^{* 2}}+\boldsymbol{\frac{\tau_{\rm m}}{L_{\rm{cell}}^2 / D_0}} \frac{\partial^2 Y_i^*}{\partial y^{* 2}}+\boldsymbol{\frac{\tau_{\rm m}}{H_{\rm{ch}}^2 / D_0}} \frac{\partial^2 Y_i^*}{\partial z^{* 2}}\right)=0$                   & Fluid channel \\
$\varepsilon \frac{\partial Y_i^*}{\partial t^*}+\boldsymbol{\frac{\tau_{\rm m}}{L_{\rm{cell}} / V_{\rm in}}} \vec{V}^* \cdot \nabla^* Y_i^*-D_{\rm eff}^*\left(\boldsymbol{\frac{\tau_{\rm m}}{W_{\rm{cell}}^2 / D_{\rm{eff},0}}} \frac{\partial^2 Y_i^*}{\partial x^{* 2}}+\boldsymbol{\frac{\tau_{\rm m}}{L_{\rm{cell}}^2 / D_{\rm{eff},0}}} \frac{\partial^2 Y_i^*}{\partial y^{* 2}}+\boldsymbol{\frac{\tau_{\rm m}}{\delta_{\rm{DL}}^2 / D_{\rm{eff},0}}} \frac{\partial^2 Y_i^*}{\partial z^{* 2}}\right)=0$                            & ADL,CDL        \\
$\varepsilon \frac{\partial Y_i^*}{\partial t^*}+\boldsymbol{\frac{\tau_{\rm m}}{L_{\rm{cell }} / V_{\rm in}}} \vec{V}^* \cdot \nabla^* Y_i^*- D_{\rm eff}^*\left(\boldsymbol{\frac{\tau_{\rm m}}{W_{\rm{cell }}^2 / D_{\rm{eff},0}}} \frac{\partial^2 Y_i^*}{\partial x^{* 2}}+\boldsymbol{\frac{\tau_{\rm m}}{L_{\rm{cell }}^2 / D_{\rm{eff},0}}} \frac{\partial^2 Y_i^*}{\partial y^{* 2}}+\boldsymbol{\frac{\tau_{\rm m}}{\delta_{\rm{FL}}^2 / D_{\rm{eff},0}}} \frac{\partial^2 Y_i^*}{\partial z^{* 2}}\right)=\frac{S_{{\rm m},i}^*}{\rho^*}$                            & AFL,CFL        \\
\textit {Conservation of energy}  &                \\
$\frac{\partial T^*}{\partial t^*}+\boldsymbol{\frac{\tau_{\rm h}}{L_{\rm{cell }} / V_{\rm in}}} \vec{V}^* \cdot \nabla^* T^*-\alpha^*\left(\boldsymbol{\frac{\tau_{\rm h}}{W_{\rm{cell}}^2 / \alpha_0}} \frac{\partial^2 T^*}{\partial x^{* 2}}+\boldsymbol{\frac{\tau_{\rm h}}{L_{\rm{cell}}^2 / \alpha_0}} \frac{\partial^2 T^*}{\partial y^{* 2}}+\boldsymbol{\frac{\tau_{\rm h}}{H_{\rm{ch}}^2 / \alpha_0}} \frac{\partial^2 T^*}{\partial z^{* 2}}\right)=0$                            & Fluid channel \\
$\frac{\partial T^*}{\partial t^*}+\boldsymbol{\frac{\tau_{\rm h}}{L_{\rm{cell }} / V_{\rm in}}} \frac{\rho^* c_{p}^*}{\rho_{\rm{eff }}^* c_{p,\rm{eff }}^*}\vec{V}^* \cdot \nabla^* T^*-\alpha_{\rm{eff }}^*\left(\boldsymbol{\frac{\tau_{\rm h}}{W_{\rm{cell }}^2 / \alpha_{\rm{eff,0 }}}} \frac{\partial^2 T^*}{\partial x^{* 2}}+\boldsymbol{\frac{\tau_{\rm h}}{L_{\rm{cell }}^2 / \alpha_{\rm{eff,0 }}}} \frac{\partial^2 T^*}{\partial y^{* 2}}+\boldsymbol{\frac{\tau_{\rm h}}{\delta_{\rm{DL}}^2 / \alpha_{\rm{eff,0 }}}} \frac{\partial^2 T^*}{\partial z^{* 2}}\right)=\frac{S_{\rm{h}}^*}{\rho_{\rm{eff }}^* c_{p,\rm{eff }}^*}$                            & ADL, CDL         \\
$\frac{\partial T^*}{\partial t^*}+\boldsymbol{\frac{\tau_{\rm h}}{L_{\rm{cell }} / V_{\rm in}}} \frac{\rho^* c_{p}^*}{\rho_{\rm{eff }}^* c_{p,\rm{eff }}^*} \vec{V}^* \cdot \nabla^* T^*-\alpha_{\rm{eff }}^*\left(\boldsymbol{\frac{\tau_{\rm h}}{W_{\rm{cell }}^2 / \alpha_{\rm{eff,0 }}}} \frac{\partial^2 T^*}{\partial x^{* 2}}+\boldsymbol{\frac{\tau_{\rm h }}{L_{\rm{cell }}^2 / \alpha_{\rm{eff,0 }}}} \frac{\partial^2 T^*}{\partial y^{* 2}}+\boldsymbol{\frac{\tau_{\rm h}}{\delta_{\rm{FL}}^2 / \alpha_{\rm{eff,0 }}}} \frac{\partial^2 T^*}{\partial z^{* 2}}\right)=\frac{S_{\rm{h}}^*}{\rho_{\rm{eff }}^* c_{p,\rm{eff }}^*}$                            & AFL, CFL         \\
$\frac{\partial T^*}{\partial t^*}-\alpha_{\rm{s}}^*\left(\boldsymbol{\frac{\tau_{\rm h}}{W_{\rm{cell }}^2 / \alpha_{\rm s,0}}} \frac{\partial^2 T^*}{\partial x^{* 2}}+\boldsymbol{\frac{\tau_{\rm h}}{L_{\rm{cell }}^2 / \alpha_{\rm s,0}}} \frac{\partial^2 T^*}{\partial y^{* 2}}+\boldsymbol{\frac{\tau_{\rm h}}{\delta_{\rm{s }}^2 / \alpha_{\rm s,0}}} \frac{\partial^2 T^*}{\partial z^{* 2}}\right)=\frac{S_{\rm{h}}^*}{\rho_{\rm{s}}^* c_{p,\rm{s}}^*}$                            & Solid \tnote{3}   \\ \hline
\end{tabular}
\begin{tablenotes}
\footnotesize
\item[1] Porous media indicates ADL, AFL, CDL, and CFL.
\item[2] Solid domain includes interconnect and solid oxide electrolyte.
\item[3] Subscript `0' denotes the reference values for fluid and material properties, which are determined under inlet temperature $T_{\rm in}$, pressure $P_{\rm in}$ and species mole composition $X_{\rm in}^{\rm H_2O/O_2/H_2/N_2}$. The subscript `eff' denotes the effective properties of porous media, which are calculated by volumetrically averaging the fluid and solid properties \cite{LIANG2023116759}. For example, the effective thermal conductivity of porous media is calculated by $k_{\rm eff} = \varepsilon k + (1 - \varepsilon) k_{\rm s}$.
\end{tablenotes}
\end{threeparttable}
\end{table}

\section{Theoretical derivations of the characteristic time} \label{Sec:parametric}
To comprehensively analyze the impact of various parameters on the transient characteristics of SOC, a meticulous parametric study was conducted through numerical simulations by altering the dimensions, material properties, and operating conditions of SOC. The boundary conditions (BC) of simulation are presented in Table~\ref{Tab:BC}. The parametric study consists of two parts, one concentrating on the transient mass-transfer behavior and the other on the thermal behavior. The simulation data was analyzed to determine the characteristic times of heat and mass transfer in SOC.

\begin{table}[]
\centering
\caption{Boundary conditions for the CFD simulation. The operating pressure is $p_0=1\,{\rm atm}$.}
\label{Tab:BC}
\small
\begin{threeparttable}
\begin{tabular}{@{}lllllll@{}}
\hline
 &
  \begin{tabular}[c]{@{}l@{}}Inlet \tnote{1} \end{tabular} &
  \begin{tabular}[c]{@{}l@{}}Outlet \end{tabular} &
  \begin{tabular}[c]{@{}l@{}}$x=0$ ,\\ $x/W_{\rm cell} = 1$\end{tabular} &
  \begin{tabular}[c]{@{}l@{}}$z=0$ \end{tabular} &
  \begin{tabular}[c]{@{}l@{}}$z/\delta_{\rm cell}=1$ \end{tabular} &
  \begin{tabular}[c]{@{}l@{}}Other \\ surfaces\end{tabular} \\ \hline
Momentum   & \begin{tabular}[c]{@{}l@{}} $V_{\rm in}^{\rm air}$, $V_{\rm in}^{\rm fuel}$ \end{tabular} & $p_{\rm gauge}=0$ & Zero flux &     N.A.             &         N.A.           & $V = 0$     \\
Thermal    & $T_{\rm in}$      &    N.A.   & Zero flux & Zero flux          & Zero flux            & Zero flux \\
Species    & \makecell[l]{$X_{\rm in}^{\rm O_2}=0.2$, $X_{\rm in}^{\rm N_2}=0.8$, $X_{\rm in}^{\rm H_2O}$, $X_{\rm in}^{\rm H_2}$}                                 &   N.A.    & Zero flux &       N.A.      &    N.A.       & Zero flux \\
Electrical &   N.A.  &    N.A.   & Zero flux & $\phi_{\rm ele} = 0$ & Specified flux \tnote{2} & Zero flux \\ \hline
\end{tabular}
\begin{tablenotes}
	\footnotesize
 	\item[1] The model encompasses two inlets: the fuel channel inlet and the air channel inlet. The fuel inlet solely comprises H$_2$O and H$_2$, while the air inlet solely comprises O$_2$ and N$_2$. In the simulation, the two inlets may exhibit different velocities but maintain an identical temperature.
	\item[2] The electric current $i$ on the surface is a time-dependent function. $i$ initiates at $i_{t=0}$ and transitions to $i_{t>0}$ in a very short time, and then maintains at $i_{t>0}$.
 	\item[ ] N.A. means `not applicable'. $V_{\rm in}$, $T_{\rm in}$, $X_{\rm H_2O}$, $X_{\rm H_2}$, $i_{t=0}$, and $i_{t>0}$ are manipulated variables in the parametric study, while the other boundary conditions are controlled. 
\end{tablenotes}
\end{threeparttable}
\end{table}

\subsection{Characteristic time of mass transfer}
The transport rate of reactants and products to or from FL significantly influences the electrochemical reaction rate and dynamic response of SOC \cite{BAE2018405}. The first part of the parametric study is to understand the transient mass-transfer behavior of SOC (including both SOEC and SOFC). We manipulated the parameters Parameters that directly impact mass transfer were manipulated. Table\,\ref{Tab:ParaMass} shows the base case of the parametric study and the manipulated variables. In runs 1 to 23, the cell operates under SOEC mode, while the remaining runs operate under SOFC mode. All the runs are following the same simulation procedure but with different manipulated variables. Each case commences at its steady state with $i_{t=0}$. Then, the current applied on the SOC changes from $i_{t=0}$ to $i_{t>0}$ in a very small time step ($<10^{-5}$\,s) and is held at $i_{t>0}$ until the end of simulation. The electrical behaviors and average mole fractions of species in FL are monitored throughout the simulation. Since the current directly reflects the generation or consumption of species in FL, the response of species can be interpreted as the gaseous response after a step change of mass source. In addition, temperature variation exerts a negligible impact on the step response of gas within the simulation time ($<1$\,s) due to the considerably faster response of gas compared to heat \cite{LIANG2023116759}. To further exclude the effect of temperature variation on mass transfer, heat sources of SOC were set to zero during simulation for gaseous response. 

\begin{table}[]
\centering
\caption{Table of manipulated variables used in the parametric study of mass transfer. A `–’ indicates that the parameter is consistent with the base case.}
\label{Tab:ParaMass}
%\footnotesize
\begin{threeparttable}
\resizebox{\linewidth}{23cm}{
\rotatebox{90}{
\begin{tabular}{lllllllllllllllllllll}
\hline
Name                                  & Run & $\delta_{\rm FL}^{\rm air}$ & $\delta_{\rm FL}^{\rm fuel}$ & $\delta_{\rm DL}^{\rm fuel}$ & $\delta_{\rm DL}^{\rm air}$ & $H_{\rm ch}$ & $W_{\rm ch}$ & $W_{\rm cell}$ & $L_{\rm cell}$ & $\varepsilon_{\rm DL}^{\rm fuel}$ & $\varepsilon_{\rm FL}^{\rm fuel}$ & $\varepsilon_{\rm DL}^{\rm air}$ & $\varepsilon_{\rm FL}^{\rm air}$ & $i_{t=0}$      & $i_{t>0}$  & $T_{\rm in}$ & $X_{\rm in}^{\rm H_2O}$ & $X_{\rm in}^{\rm H_2}$ & $V_{\rm in}^{\rm air}$ & $V_{\rm in}^{\rm fuel}$ \\
                                      &     & [m]                     & [m]                      & [m]                      & [m]                     & [m]      & [m]      & [m]        & [m]        &                                   &                                   &                                  &                                  & [A/cm$^2$] & [A/cm$^2$] & [K]      &                         &                        & [m/s]              & [m/s]               \\ \hline
Base                                  & 1   & 7.0E-06                     & 7.0E-06                      & 1.0E-03                      & 4.5E-05                     & 1.0E-03      & 1.0E-03      & 2.0E-03        & 0.10           & 0.38                              & 0.20                              & 0.27                             & 0.27                             & -1.00          & -0.50          & 1073.15      & 0.70                    & 0.30                   & 2.00                   & 2.00                    \\
$\delta$1                             & 2   & 1.4E-05                     & 1.4E-05                      & --                           & --                          & --           & --           & --             & --             & --                                & --                                & --                               & --                               & --             & --             & --           & --                      & --                     & --                     & --                      \\
$\delta$2                             & 3   & --                          & --                           & 5.0E-04                      & --                          & --           & --           & --             & --             & --                                & --                                & --                               & --                               & --             & --             & --           & --                      & --                     & --                     & --                      \\
$\delta$3                             & 4   & --                          & --                           & --                           & 9.0E-05                     & --           & --           & --             & --             & --                                & --                                & --                               & --                               & --             & --             & --           & --                      & --                     & --                     & --                      \\
H1                                    & 5   & --                          & --                           & --                           & --                          & 2.0E-03      & --           & --             & --             & --                                & --                                & --                               & --                               & --             & --             & --           & --                      & --                     & --                     & --                      \\
W1                                    & 6   & --                          & --                           & --                           & --                          & --           & 5.0E-04      & 1.0E-03        & --             & --                                & --                                & --                               & --                               & --             & --             & --           & --                      & --                     & --                     & --                      \\
L1                                    & 7   & --                          & --                           & --                           & --                          & --           & --           & --             & 0.05           & --                                & --                                & --                               & --                               & --             & --             & --           & --                      & --                     & --                     & --                      \\
{\color[HTML]{333333} $\varepsilon$1} & 8   & --                          & --                           & --                           & --                          & --           & --           & --             & --             & 0.76                              & --                                & --                               & --                               & --             & --             & --           & --                      & --                     & --                     & --                      \\
{\color[HTML]{333333} $\varepsilon$2} & 9   & --                          & --                           & --                           & --                          & --           & --           & --             & --             & --                                & 0.40                              & --                               & --                               & --             & --             & --           & --                      & --                     & --                     & --                      \\
{\color[HTML]{333333} $\varepsilon$3} & 10  & --                          & --                           & --                           & --                          & --           & --           & --             & --             & --                                & --                                & 0.54                             & --                               & --             & --             & --           & --                      & --                     & --                     & --                      \\
{\color[HTML]{333333} $\varepsilon$4} & 11  & --                          & --                           & --                           & --                          & --           & --           & --             & --             & --                                & --                                & --                               & 0.54                             & --             & --             & --           & --                      & --                     & --                     & --                      \\
I1                                    & 12  & --                          & --                           & --                           & --                          & --           & --           & --             & --             & --                                & --                                & --                               & --                               & --             & -0.75          & --           & --                      & --                     & --                     & --                      \\
I2                                    & 13  & --                          & --                           & --                           & --                          & --           & --           & --             & --             & --                                & --                                & --                               & --                               & --             & -1.25          & --           & --                      & --                     & --                     & --                      \\
T1                                    & 14  & --                          & --                           & --                           & --                          & --           & --           & --             & --             & --                                & --                                & --                               & --                               & --             & --             & 1023.15      & --                      & --                     & --                     & --                      \\
T2                                    & 15  & --                          & --                           & --                           & --                          & --           & --           & --             & --             & --                                & --                                & --                               & --                               & --             & --             & 1123.15      & --                      & --                     & --                     & --                      \\
X1                                    & 16  & --                          & --                           & --                           & --                          & --           & --           & --             & --             & --                                & --                                & --                               & --                               & --             & --             & --           & 0.50                    & 0.50                   & --                     & --                      \\
X2                                    & 17  & --                          & --                           & --                           & --                          & --           & --           & --             & --             & --                                & --                                & --                               & --                               & --             & --             & --           & 0.90                    & 0.10                   & --                     & --                      \\
V1                                    & 18  & --                          & --                           & --                           & --                          & --           & --           & --             & --             & --                                & --                                & --                               & --                               & --             & --             & --           & --                      & --                     & 1.50                   & 1.50                    \\
V2                                    & 19  & --                          & --                           & --                           & --                          & --           & --           & --             & --             & --                                & --                                & --                               & --                               & --             & --             & --           & --                      & --                     & 2.50                   & 2.50                    \\
V3                                    & 20  & --                          & --                           & --                           & --                          & --           & --           & --             & --             & --                                & --                                & --                               & --                               & --             & --             & --           & --                      & --                     & 4.00                   & 4.00                    \\
VX1                                   & 21  & --                          & --                           & --                           & --                          & --           & --           & --             & --             & --                                & --                                & --                               & --                               & --             & --             & --           & 1.00                    & 0.00                   & 1.00                   & 1.00                    \\
VH1                                   & 22  & --                          & --                           & --                           & --                          & 2.0E-03      & --           & --             & --             & --                                & --                                & --                               & --                               & --             & --             & --           & --                      & --                     & 1.00                   & 1.00                    \\
VH2                                   & 23  & --                          & --                           & --                           & --                          & 5.0E-04      & --           & --             & --             & --                                & --                                & --                               & --                               & --             & --             & --           & --                      & --                     & 4.00                   & 4.00                    \\
FC                                    & 24  & --                          & --                           & --                           & --                          & --           & --           & --             & --             & --                                & --                                & --                               & --                               & 1.50           & 0.50           & --           & 0.10                    & 0.90                   & 18.00                  & 2.00                    \\
FC-T                                  & 25  & --                          & --                           & --                           & --                          & --           & --           & --             & --             & --                                & --                                & --                               & --                               & 1.50           & 0.50           & 1123.15      & 0.10                    & 0.90                   & 18.00                  & 2.00                    \\
FC-I                                  & 26  & --                          & --                           & --                           & --                          & --           & --           & --             & --             & --                                & --                                & --                               & --                               & 1.50           & 1.00           & --           & 0.10                    & 0.90                   & 18.00                  & 2.00                    \\
FC-V                                  & 27  & --                          & --                           & --                           & --                          & --           & --           & --             & --             & --                                & --                                & --                               & --                               & 1.50           & 0.50           & --           & 0.10                    & 0.90                   & 18.00                  & 4.00                      \\ \hline
\end{tabular}
}}
\end{threeparttable}
\end{table}

\subsubsection{Test under SOEC mode}
\begin{figure}[h]
    \centering
    \begin{subfigure}[b]{1\textwidth}
        \centering
        \includegraphics[height=8mm]{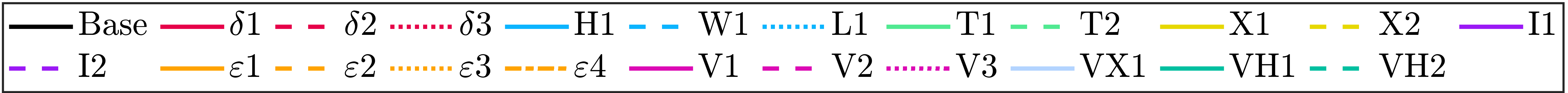}
    \end{subfigure}
    \\
    \begin{subfigure}[b]{0.34\textwidth}
        \includegraphics[width=\textwidth]{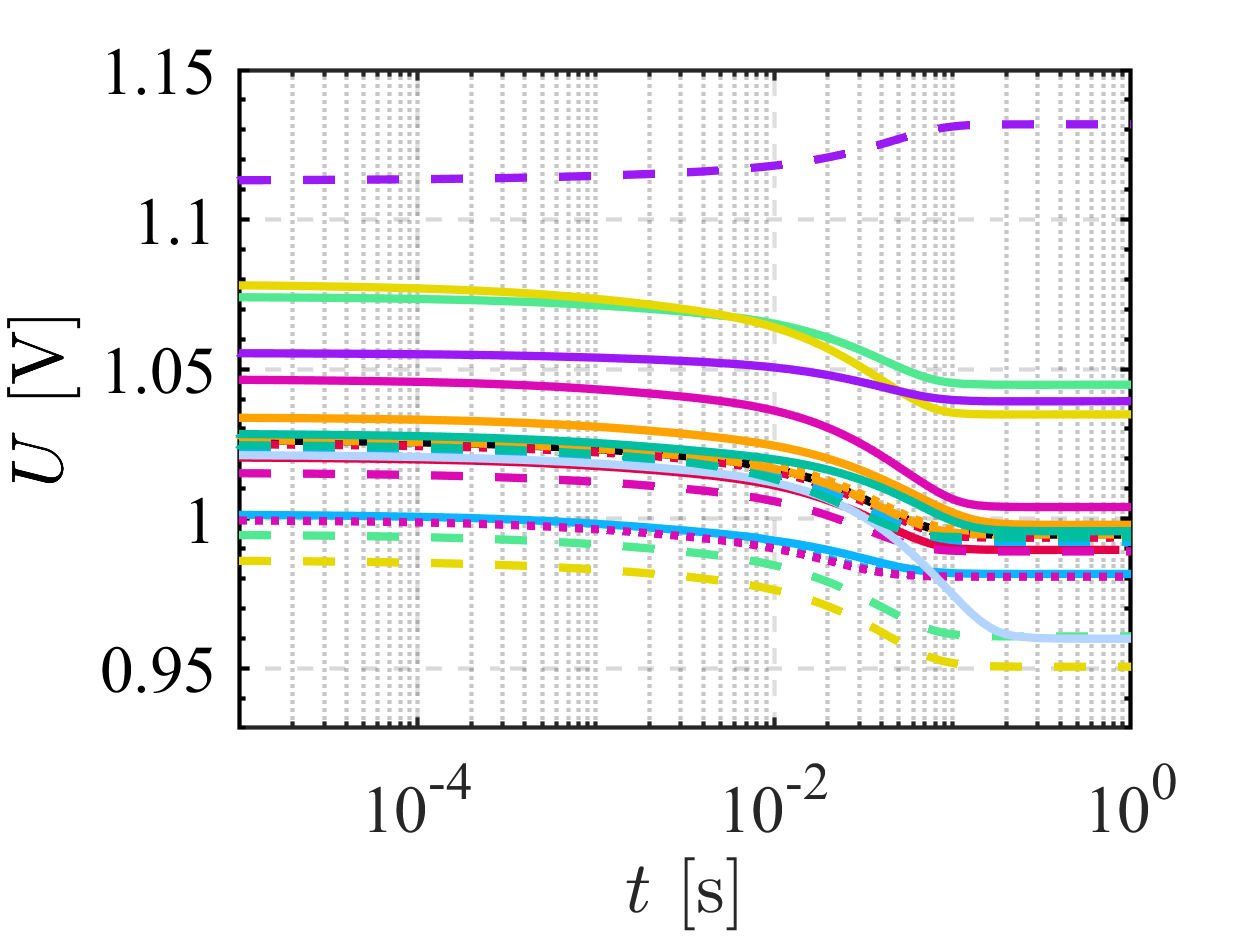}
        \caption{}
        \label{fig:SOEC_sub1}
    \end{subfigure}
    \hspace{-0.4cm}
    %\hfill
    \begin{subfigure}[b]{0.34\textwidth}
        \includegraphics[width=\textwidth]{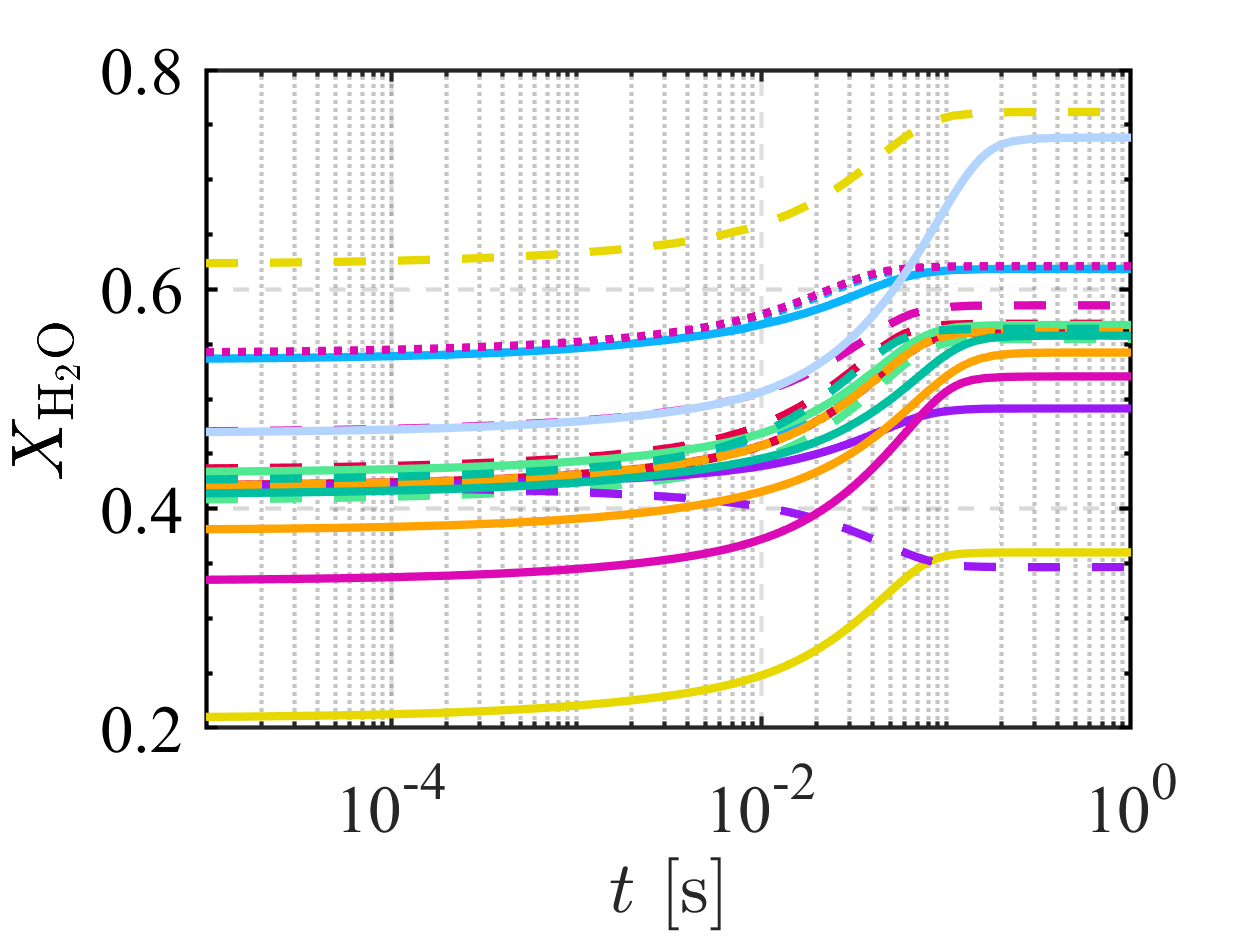}
        \caption[b]{}
        \label{fig:SOEC_sub2}
    \end{subfigure}
    \hspace{-0.4cm}
    %\hfill
    \begin{subfigure}[b]{0.34\textwidth}
        \includegraphics[width=\textwidth]{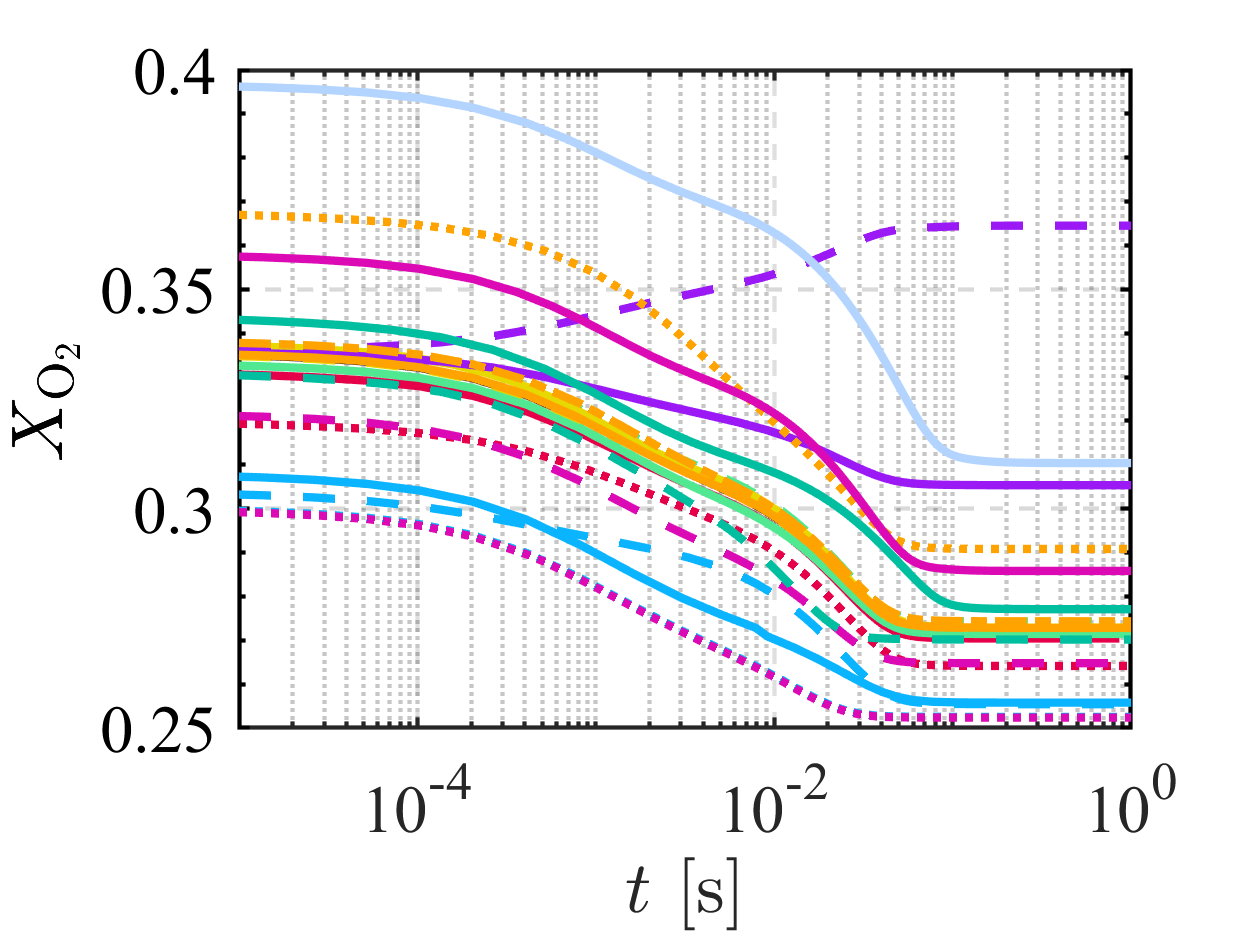}
        \caption[\vspace{-0.2cm}]{}
        \label{fig:SOEC_sub3}
    \end{subfigure}
    %\\
    \vspace{-0.15cm}
    \begin{subfigure}[b]{0.34\textwidth}
        \includegraphics[width=\textwidth]{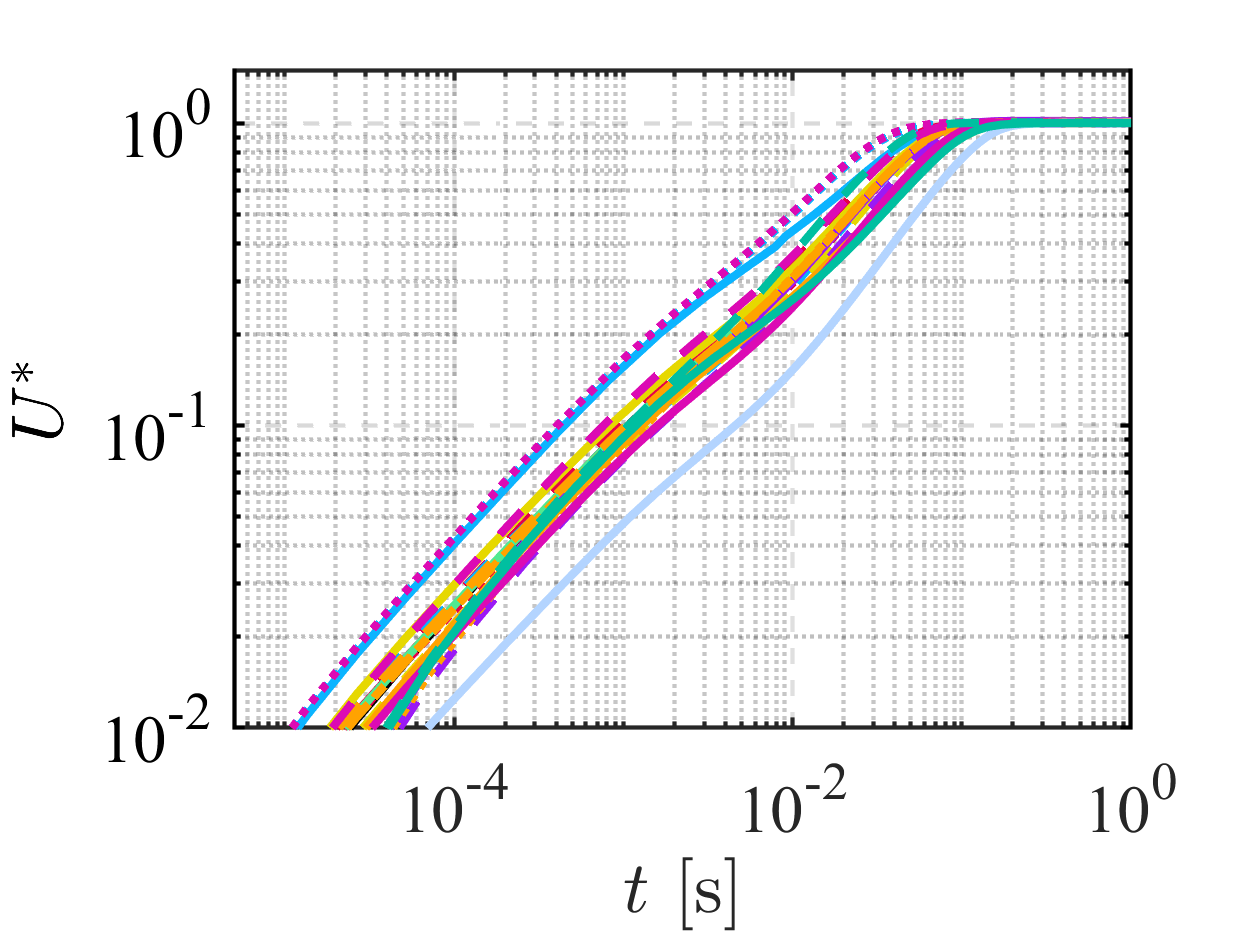}
        \caption{}
        \label{fig:SOEC_sub4}
    \end{subfigure}
    \hspace{-0.4cm}
    \begin{subfigure}[b]{0.34\textwidth}
        \includegraphics[width=\textwidth]{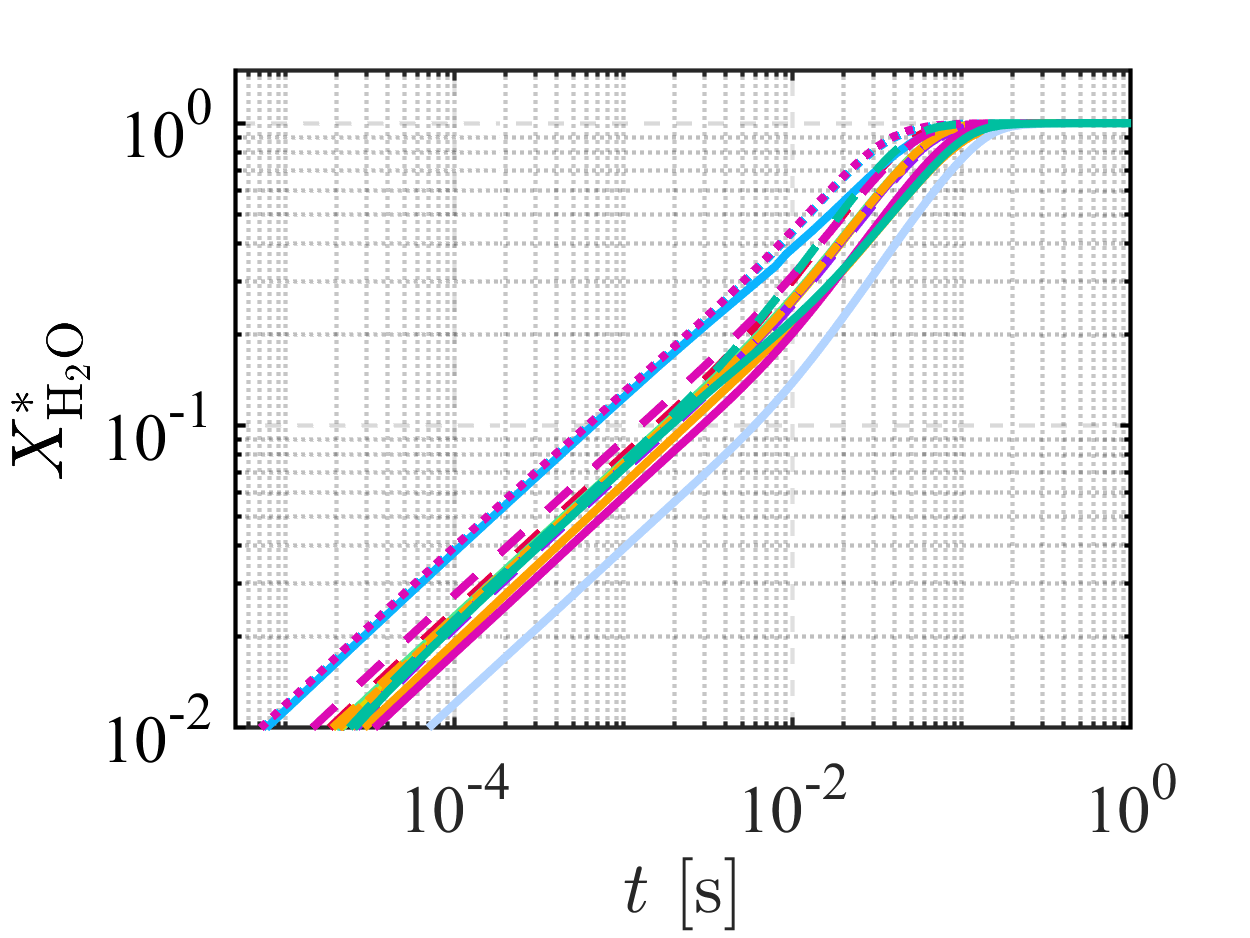}
        \caption{}
        \label{fig:SOEC_sub5}
    \end{subfigure}
    \hspace{-0.4cm}
    \begin{subfigure}[b]{0.34\textwidth}
        \includegraphics[width=\textwidth]{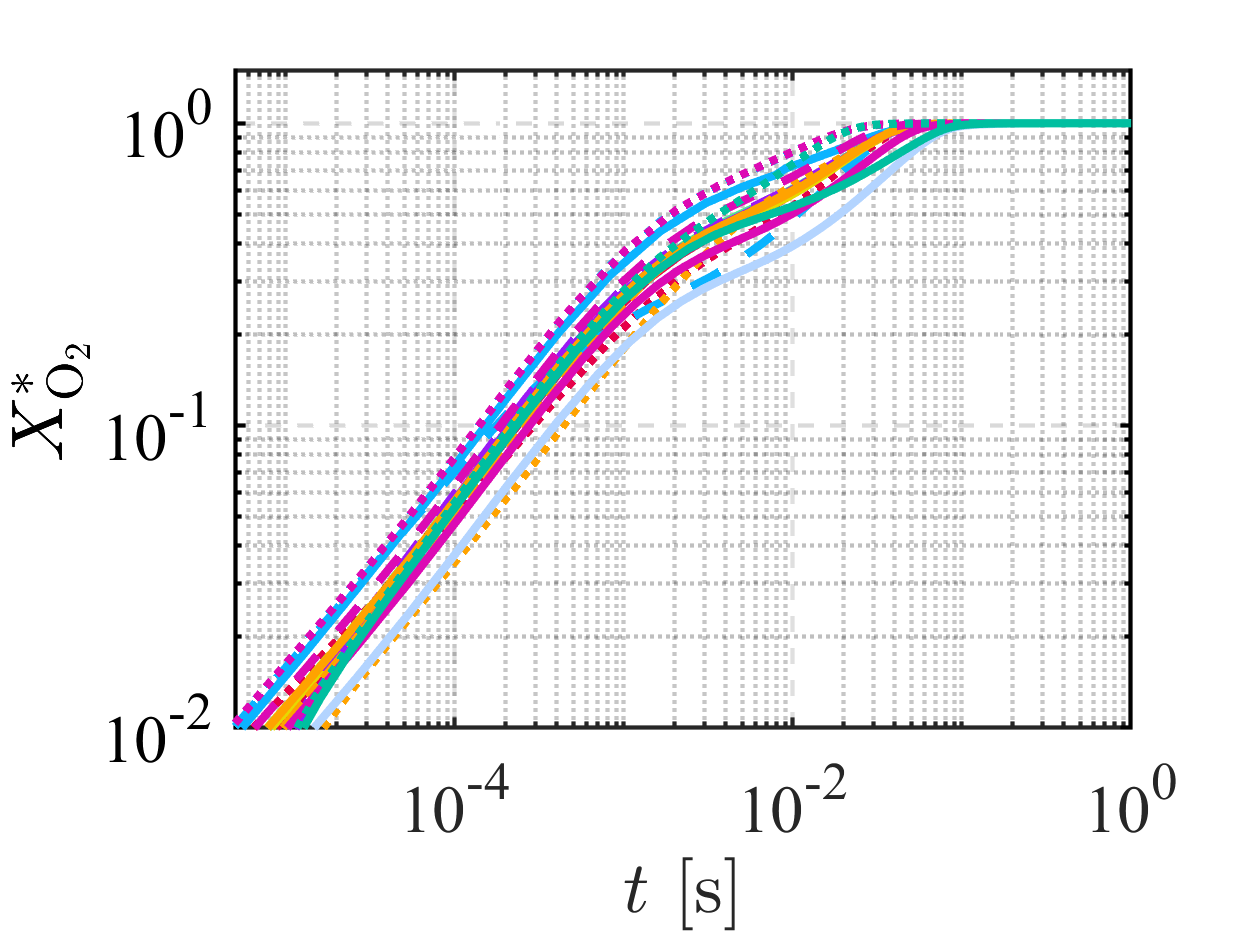}
        \caption{}
        \label{fig:SOEC_sub6}
    \end{subfigure}
    \\
    \begin{subfigure}[b]{0.34\textwidth}
        \includegraphics[width=\textwidth]{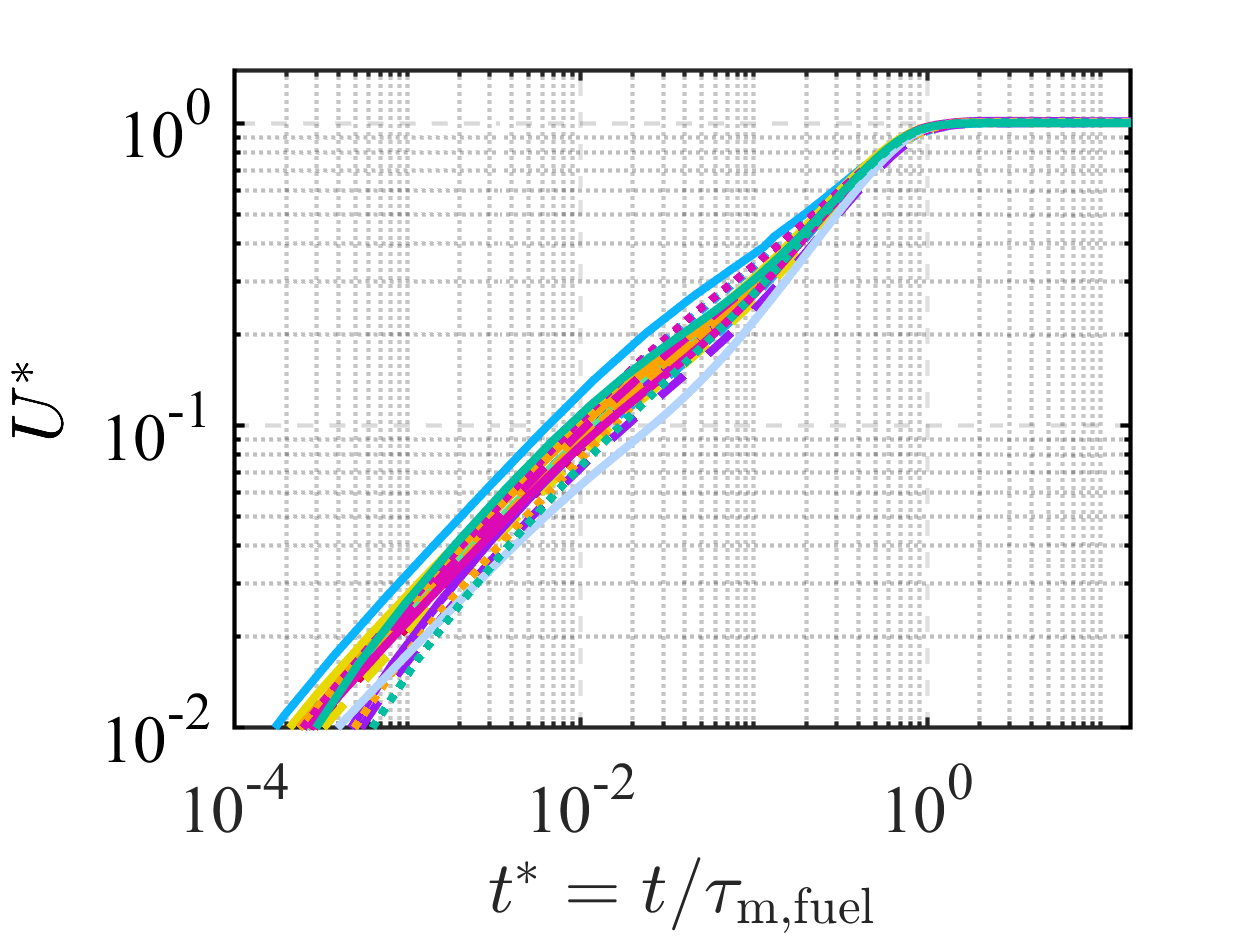}
        \caption{}
        \label{fig:SOEC_sub7}
    \end{subfigure}
    \hspace{-0.4cm}
    \begin{subfigure}[b]{0.34\textwidth}
        \includegraphics[width=\textwidth]{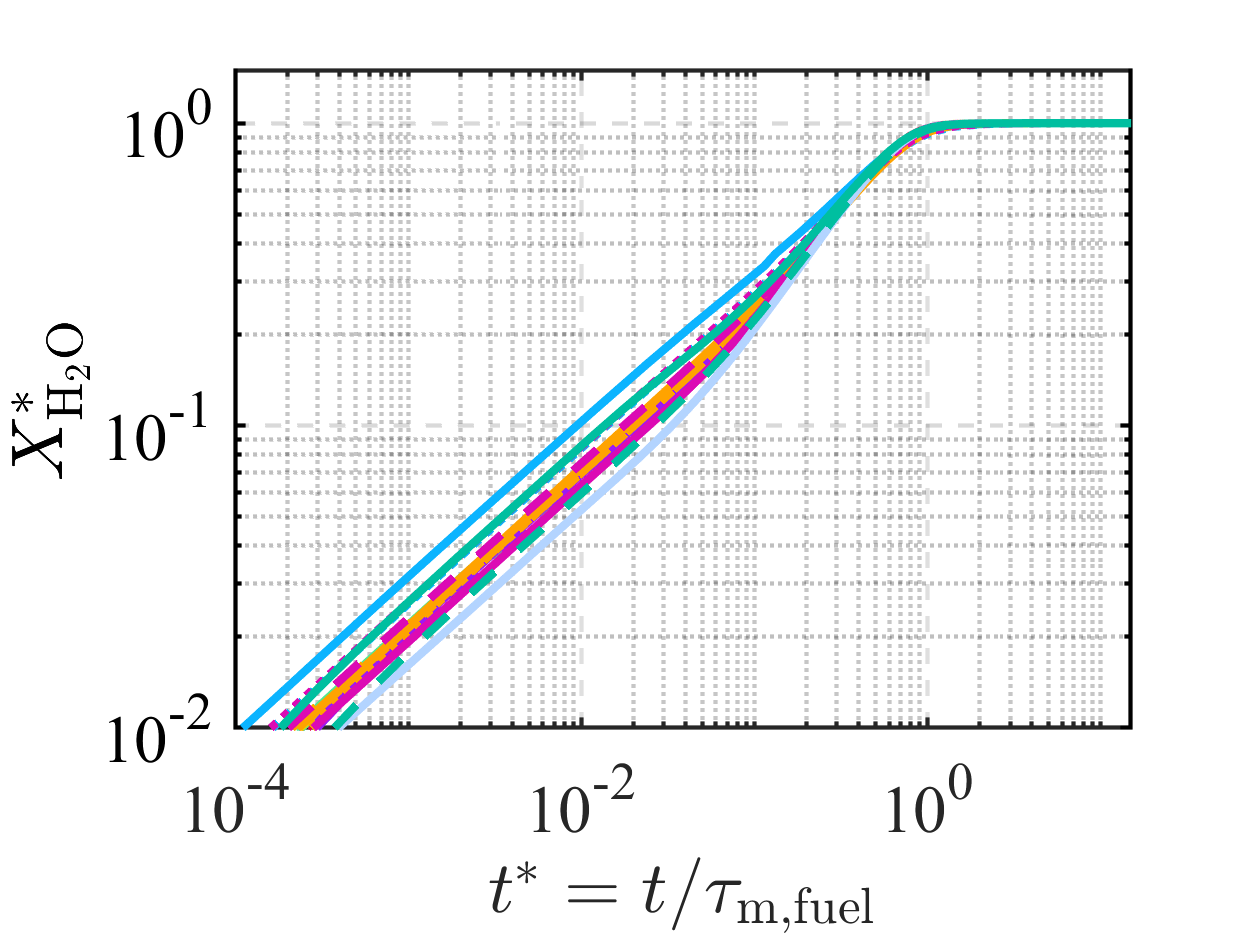}
        \caption{}
        \label{fig:SOEC_sub8}
    \end{subfigure}
    \hspace{-0.4cm}
    \begin{subfigure}[b]{0.34\textwidth}
        \includegraphics[width=\textwidth]{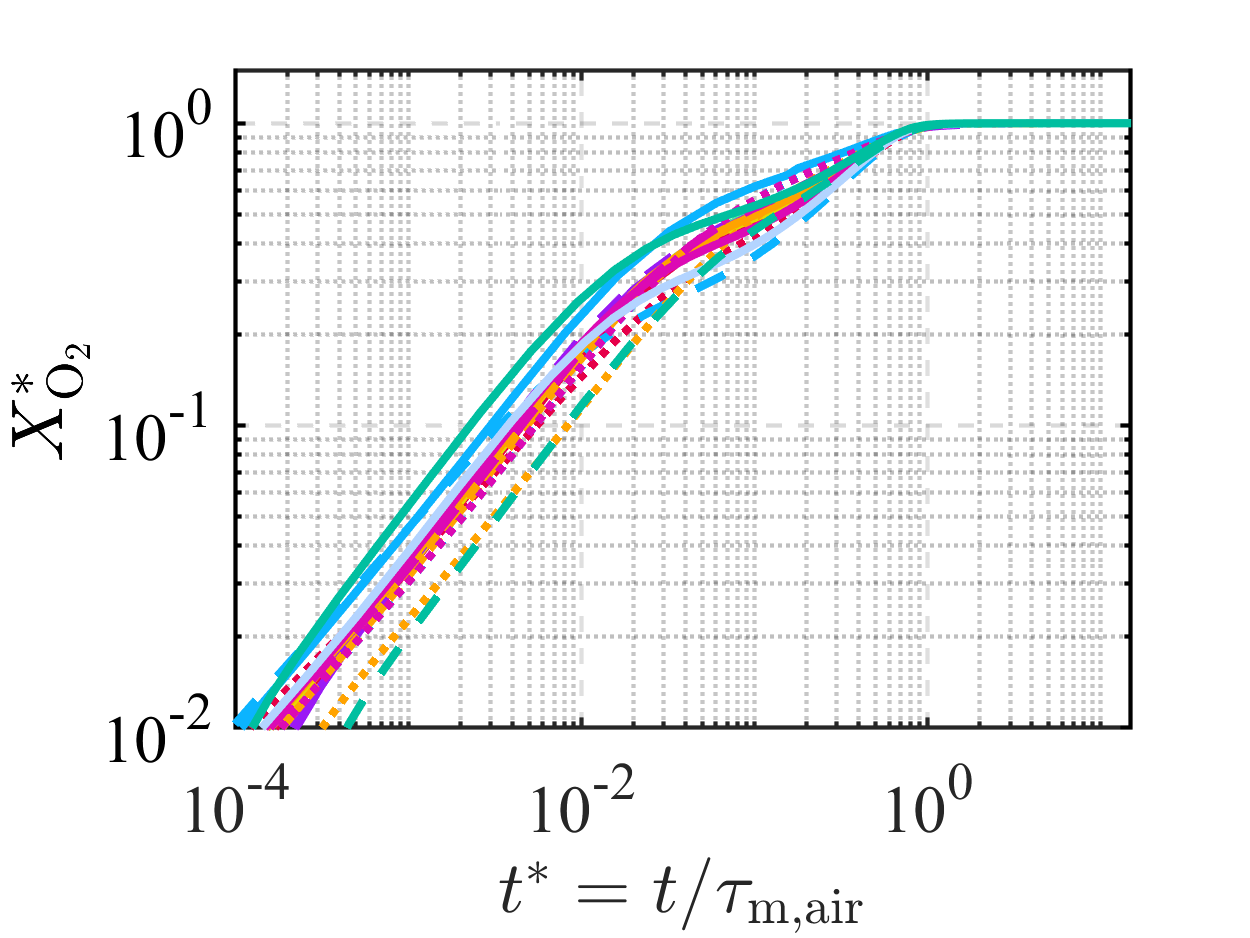}
        \caption{}
        \label{fig:SOEC_sub9}
    \end{subfigure}
    \caption{Transient responses of SOEC after step changes of current (mass source). SOEC with different dimensions, material properties, and operating conditions are compared.}
    \label{fig:SOEC_transient}
\end{figure}

The simulation results under SOEC mode are illustrated in Fig.\,\ref{fig:SOEC_transient} where, the temporal variations of voltage $U$ and mole fractions of water $X_{\rm H_2O}$ and oxygen $X_{\rm O_2}$ are presented in Fig.\,\ref{fig:SOEC_sub1}, \ref{fig:SOEC_sub2}, and \ref{fig:SOEC_sub3}. It is observed that except for case `I2' that undergoes a step increase of current density magnitude $|i|$, the remaining cases depict downward trends of $U$ and $X_{\rm O_2}$ along with upward trends of $X_{\rm H_2O}$ as results of the step-down $|i|$. This phenomenon can be attributed to the fact that the decrease in $|i|$ reduces electrolysis intensity, implying that SOEC requires less voltage $U$, consumes less H$_2$O, and produces less O$_2$. However, each case has its own unique initial and final steady state, rendering direct comparison of transients across different cases challenging.

To facilitate comparison of the relaxation time required by each case to reach its steady state, $U$, $X_{\rm H_2O}$ and $X_{\rm O_2}$ are scaled to $U^*$, $X_{\rm H_2O}^*$ and $X_{\rm O_2}^*$, respectively, as presented in Table\,\ref{Tab:Dimensional}. The temporal variations of $U^*$, $X_{\rm H_2O}^*$ and $X_{\rm O_2}^*$ are plotted in Fig.\,\ref{fig:SOEC_sub4}, \ref{fig:SOEC_sub5}, and \ref{fig:SOEC_sub6}. The values of $U^*$, $X_{\rm H_2O}^*$, and $X_{\rm O_2}^*$ represent the extent of steadiness achieved, with the ideal value of 1 indicating 100\% steady state. In this study, the time required for a response to reach 95\% steady state is considered as the relaxation time. The temporal plots of $U^*$, $X_{\rm H_2O}^*$, and $X_{\rm O_2}^*$ reveal the complicated relationship between relaxation time and SOC parameters. To elucidate this relationship, we derived the mass-transfer characteristic time $\tau_{\rm m}$ of SOCs. 

The expression for $\tau_{\rm m}$, as presented in Eq.\,(\ref{eq:tau_m}), reflects the time required for the flow to fill the void in SOC. The approximation of $\tau_{\rm m}$ is valid when the volumetric change of gas inside the SOEC can be considered negligible in comparison to the volumetric flow rate at inlets. 
\begin{equation}\label{eq:tau_m}
\tau_{\rm{m}}=\frac{\text{ Void volume }}{\text{ Volumetric flow rate }} \frac{\left[\rm{m}^3\right]}{\left[\rm{m}^3 / \rm{s}\right]} \approx 
\left\{\begin{array}{l}
\frac{W_{\rm{ch}} H_{\rm{ch}} L_{\rm{cell}}+W_{\rm{cell}} L_{\rm{cell}}\left(\varepsilon_{\rm{DL}}^{\rm{fuel}} \delta_{\rm{DL}}^{\rm{fuel}}+\varepsilon_{\rm{FL}}^{\rm{fuel}} \delta_{\rm{FL}}^{\rm{fuel}}\right)}{W_{\rm{ch}} H_{\rm{ch}} V_{\rm{in}}^{\rm{fuel}}} \quad \text{Fuel side}\\
\frac{W_{\rm{ch}} H_{\rm{ch}} L_{\rm{cell}}+W_{\rm{cell}} L_{\rm{cell}}\left(\varepsilon_{\rm{DL}}^{\rm{air}} \delta_{\rm{DL}}^{\rm{air}}+\varepsilon_{\rm{FL}}^{\rm{air}} \delta_{\rm{FL}}^{\rm{air}}\right)}{W_{\rm{ch}} H_{\rm{ch}} V_{\rm{in}}^{\rm{air}}} \quad \text{Air side}
\end{array}\right.    
\end{equation}

By scaling the time with the $\tau_{\rm m}$ of each case, Fig.\,\ref{fig:SOEC_sub7}-- \ref{fig:SOEC_sub9} are obtained, where the relaxation times of all cases converge to one point. The results show that $X_{\rm H_2O}^*$ in the fuel side and $X_{\rm O_2}^*$ in the air side take around $1 \tau_{\rm m}^{\rm fuel}$ and $1 \tau_{\rm m}^{\rm air}$, respectively, to stabilize after a step change of current. Interestingly, the relaxation time of $U^*$ is approximately  $1 \tau_{\rm m}^{\rm fuel}$, indicating that the species transport in the fuel electrode is the limiting factor of the electrical response. 

Note that, according to Eq.\,(\ref{eq:tau_m}), $\tau_{\rm m}$ may have different values on the fuel side and the air side of SOC due to the differences in flow conditions and electrode structures. This observation is consistent with Nerat's study \cite{NERAT2017728}, which claimed that the gaseous response in the fuel electrode is slower than that in the air electrode due to the significantly thicker DL in the fuel side. Furthermore, Eq.\,(\ref{eq:tau_m}) describes the relative importance of parameters, as shown in Fig\,\ref{fig:composition_void}, which compares the void volume fraction of the components of SOC. The porosity and thickness of DL and FL exert a less substantial impact on $\tau_{\rm m}$ compared to the channel due to their minor contributions to the void volume. The volumetric flow rate, as the numerator of Eq.\,(\ref{eq:tau_m}), exerts a considerable effect on $\tau_{\rm m}$. These observations are in line with the findings of Bae et al. \cite{BAE2019112152}, who reported the minor effects of microstructure and emphasized the importance of volumetric flow rate on SOFC transients.

\subsubsection{Test under SOFC mode}
\begin{figure}[h]
    \centering
    \begin{subfigure}[b]{1\textwidth}
        \centering
        \includegraphics[height=5mm]{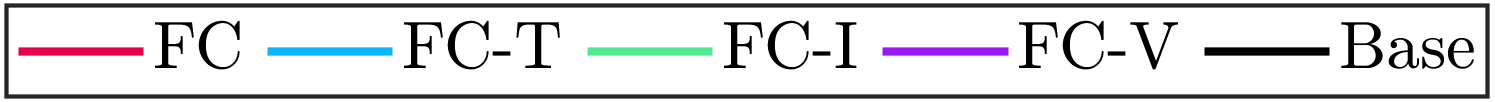}
    \end{subfigure}
    \\
    \begin{subfigure}[b]{0.33\textwidth}
        \includegraphics[width=\textwidth]{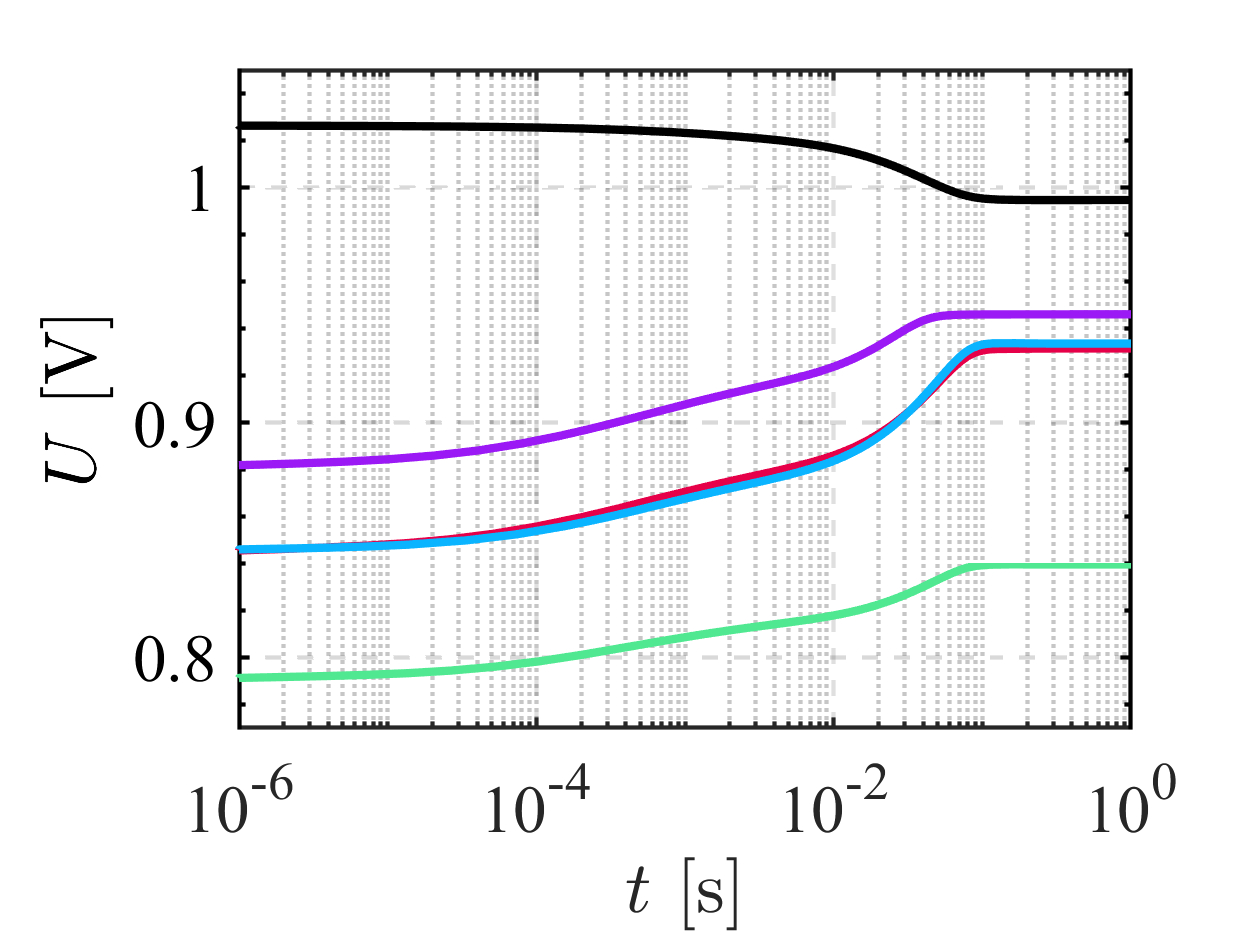}
        \caption{}
        \label{fig:SOFC_sub1}
    \end{subfigure}
    \hspace{-0.4cm}
    \begin{subfigure}[b]{0.32\textwidth}
        \includegraphics[width=\textwidth]{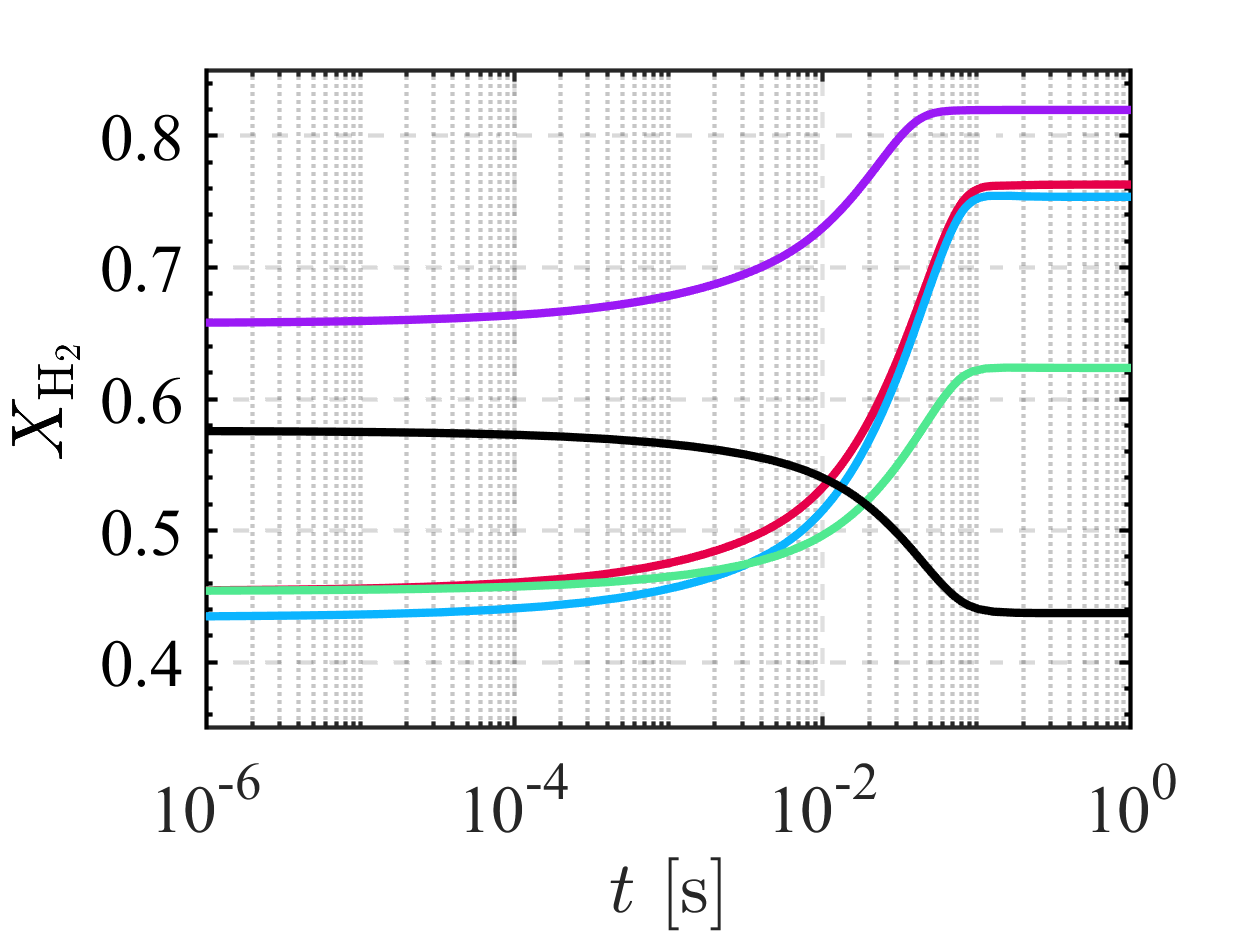}
        \caption{}
        \label{fig:SOFC_sub2}
    \end{subfigure}
    \hspace{-0.4cm}
    \begin{subfigure}[b]{0.32\textwidth}
        \includegraphics[width=\textwidth]{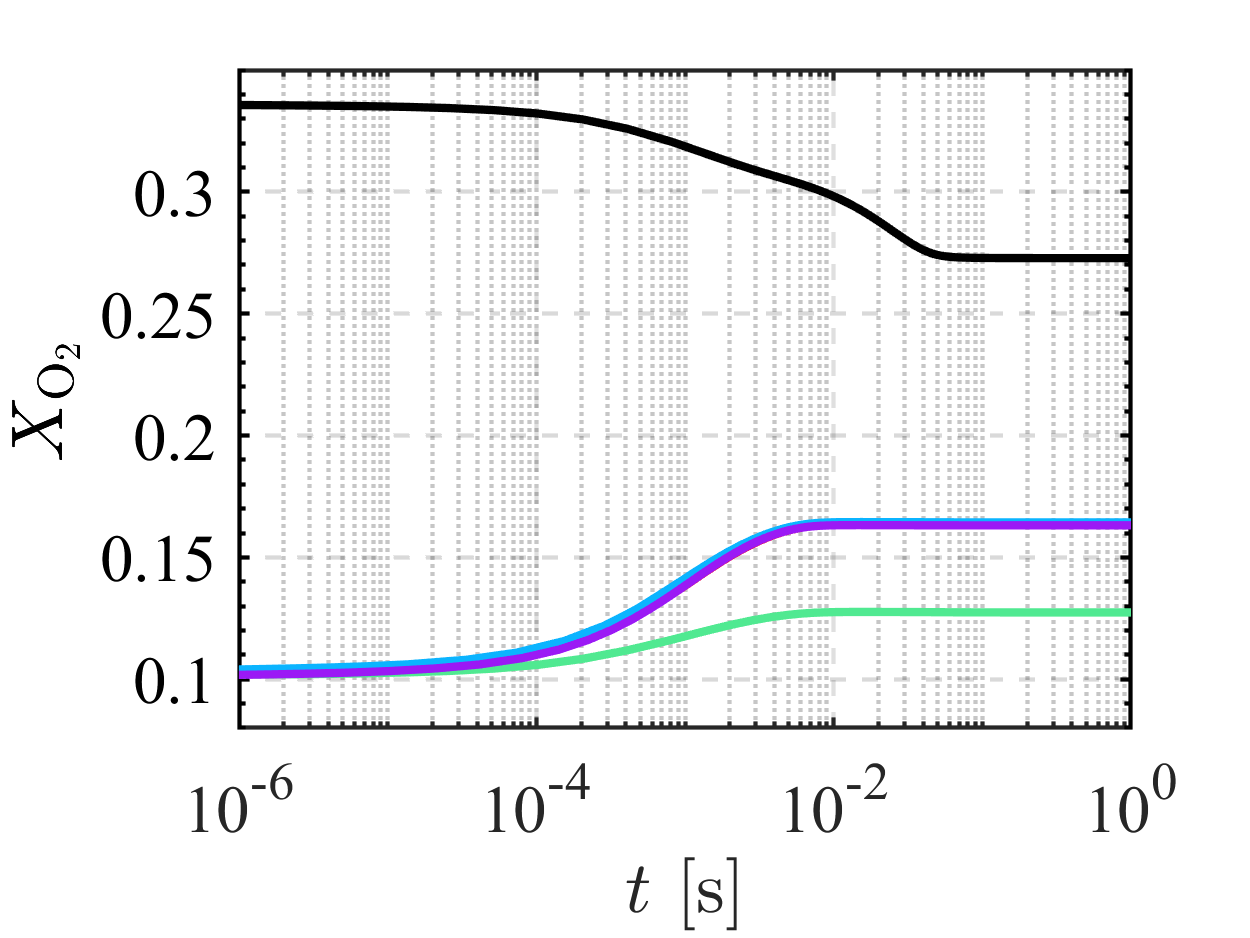}
        \caption{}
        \label{fig:SOFC_sub3}
    \end{subfigure}
    \\
    \begin{subfigure}[b]{0.32\textwidth}
        \includegraphics[width=\textwidth]{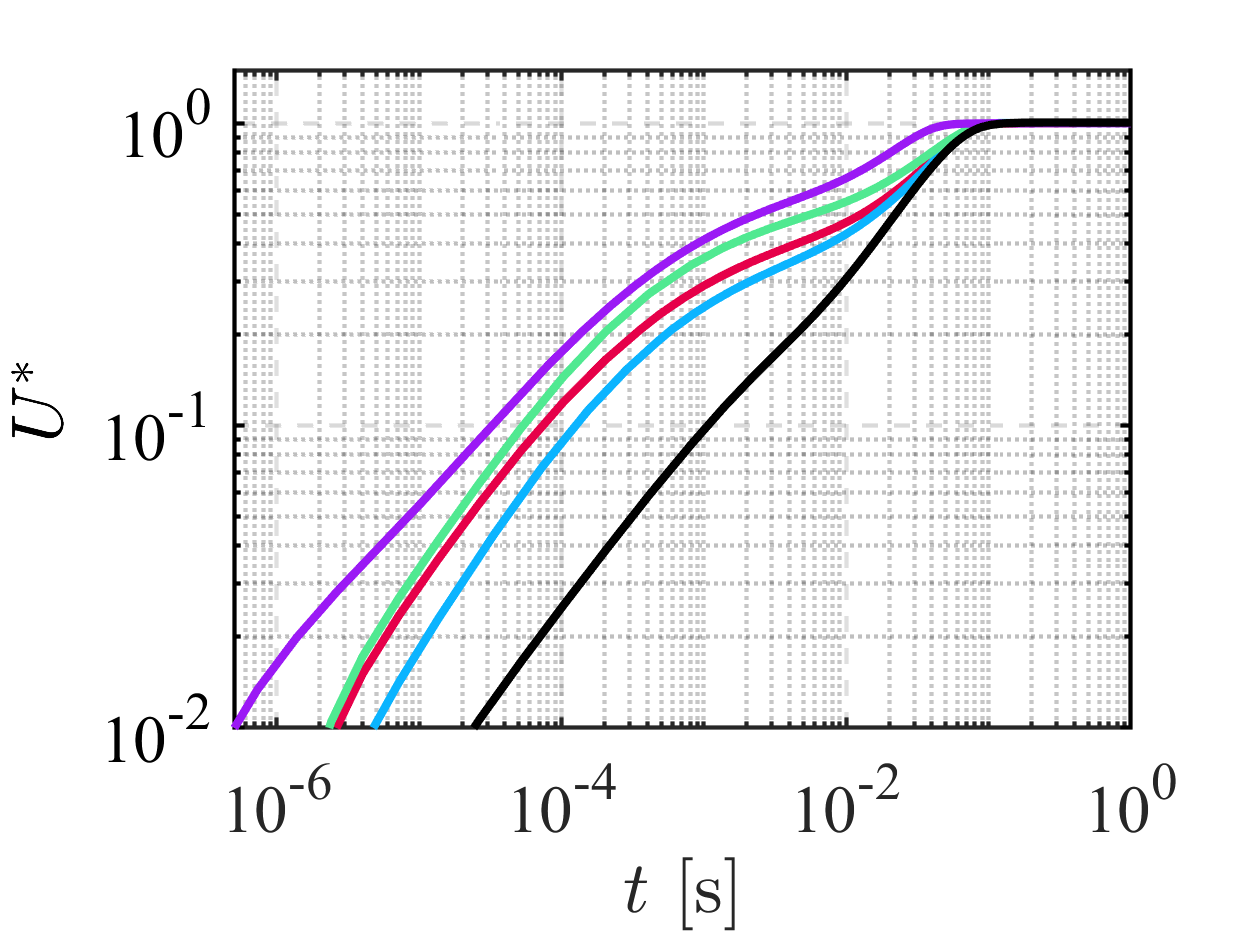}
        \caption{}
        \label{fig:SOFC_sub4}
    \end{subfigure}
    \hspace{-0.4cm}
    \begin{subfigure}[b]{0.32\textwidth}
        \includegraphics[width=\textwidth]{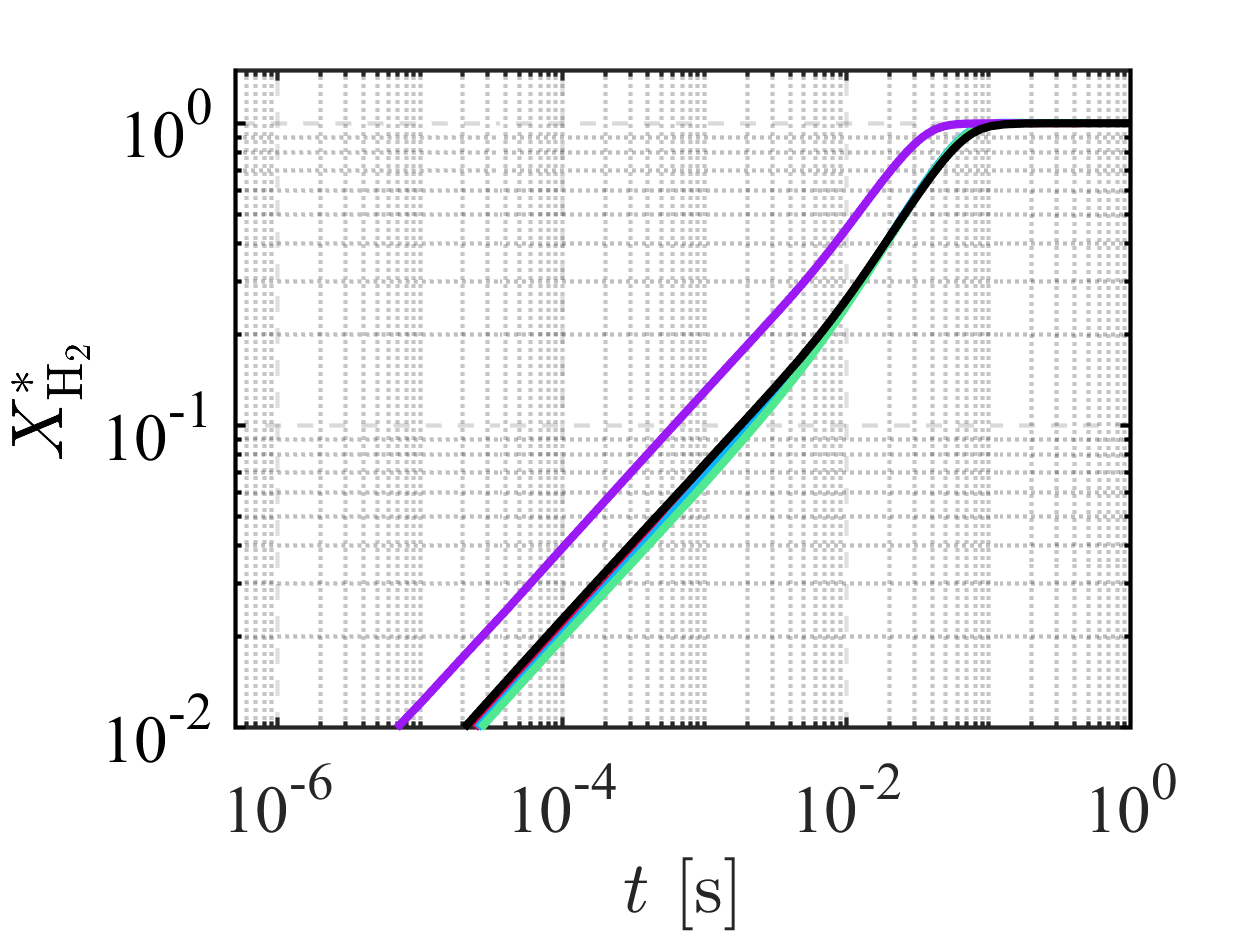}
        \caption{}
        \label{fig:SOFC_sub5}
    \end{subfigure}
    \hspace{-0.4cm}
    \begin{subfigure}[b]{0.32\textwidth}
        \includegraphics[width=\textwidth]{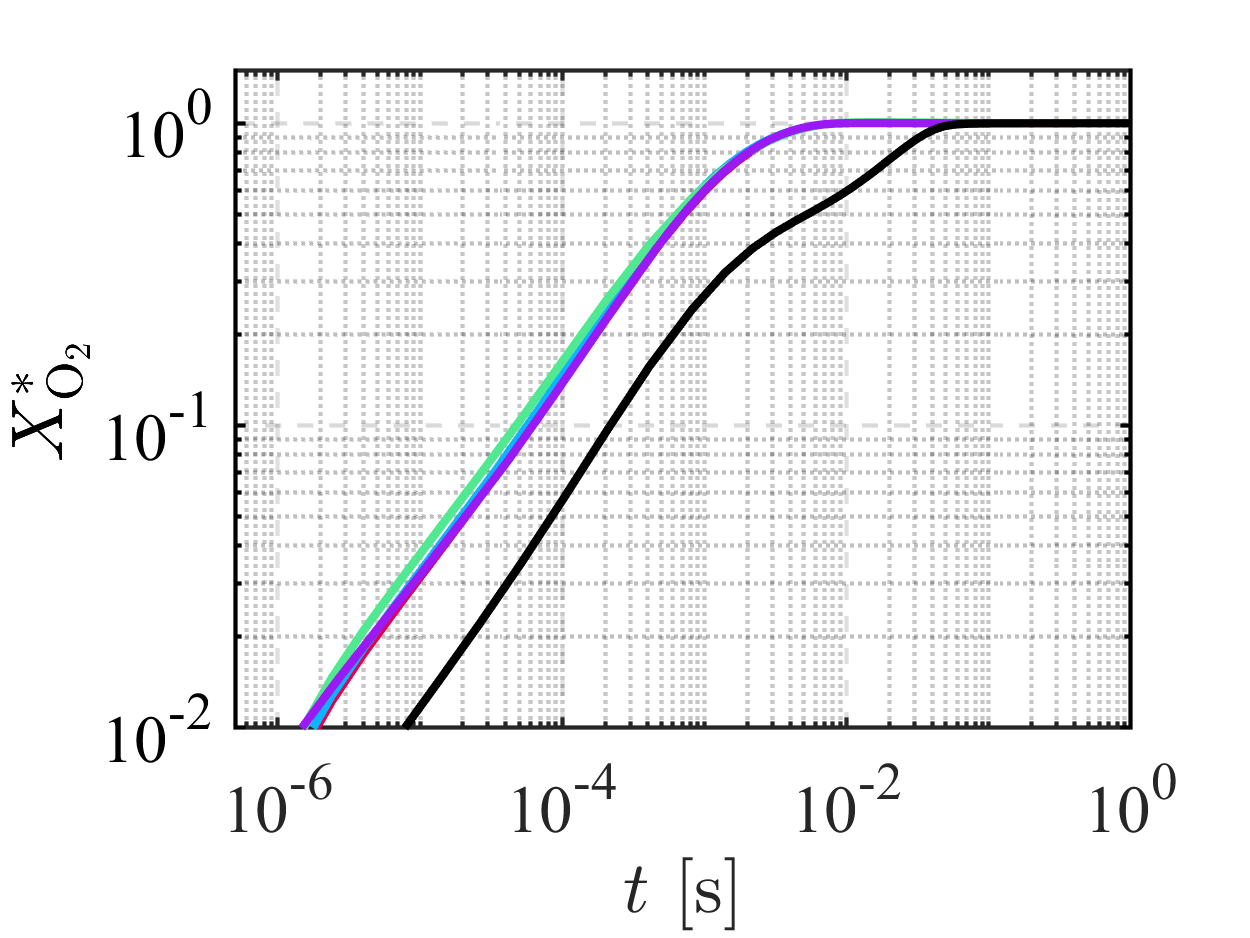}
        \caption{}
        \label{fig:SOFC_sub6}
    \end{subfigure}
    \\
    \begin{subfigure}[b]{0.32\textwidth}
        \includegraphics[width=\textwidth]{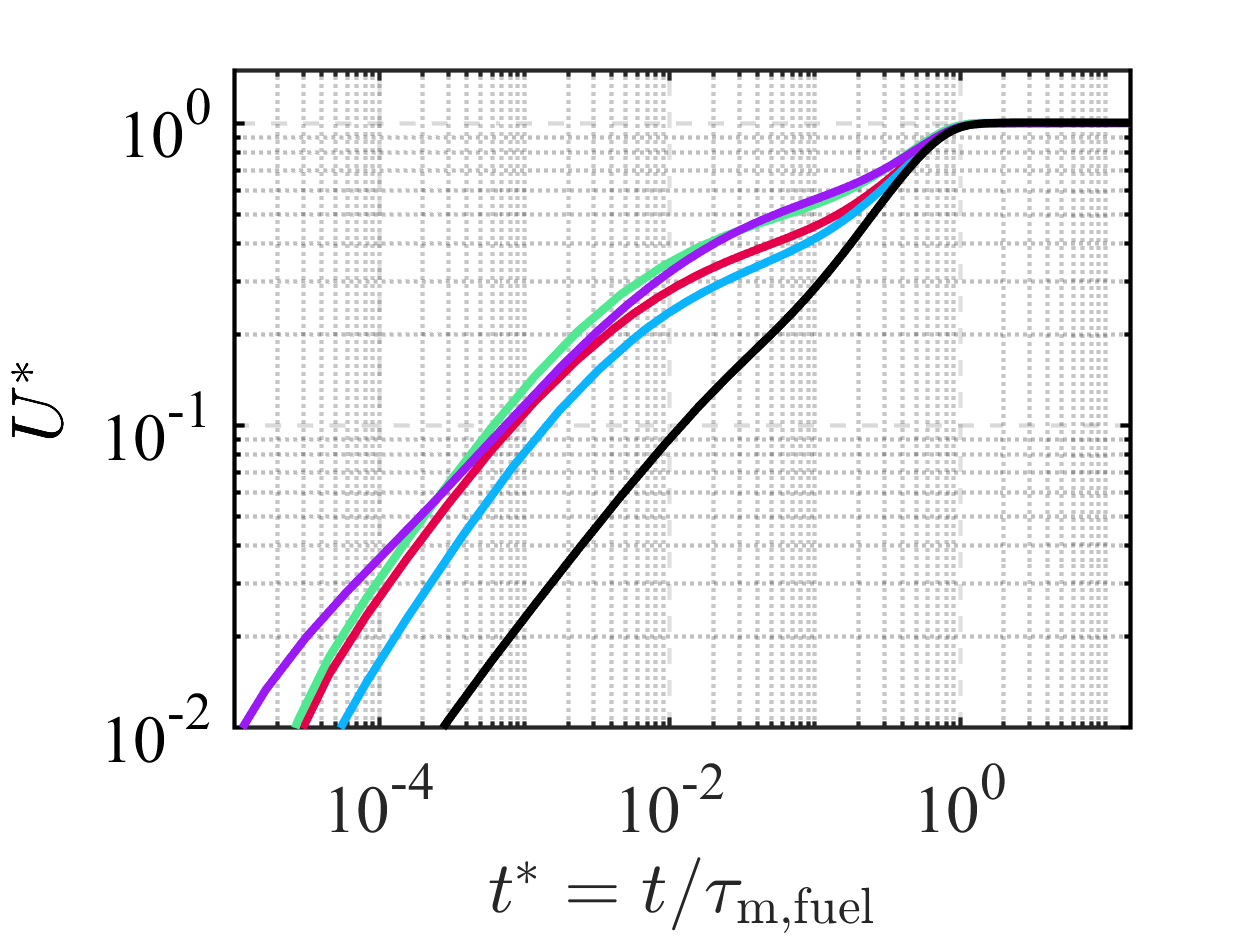}
        \caption{}
        \label{fig:SOFC_sub7}
    \end{subfigure}
    \hspace{-0.4cm}
    \begin{subfigure}[b]{0.32\textwidth}
        \includegraphics[width=\textwidth]{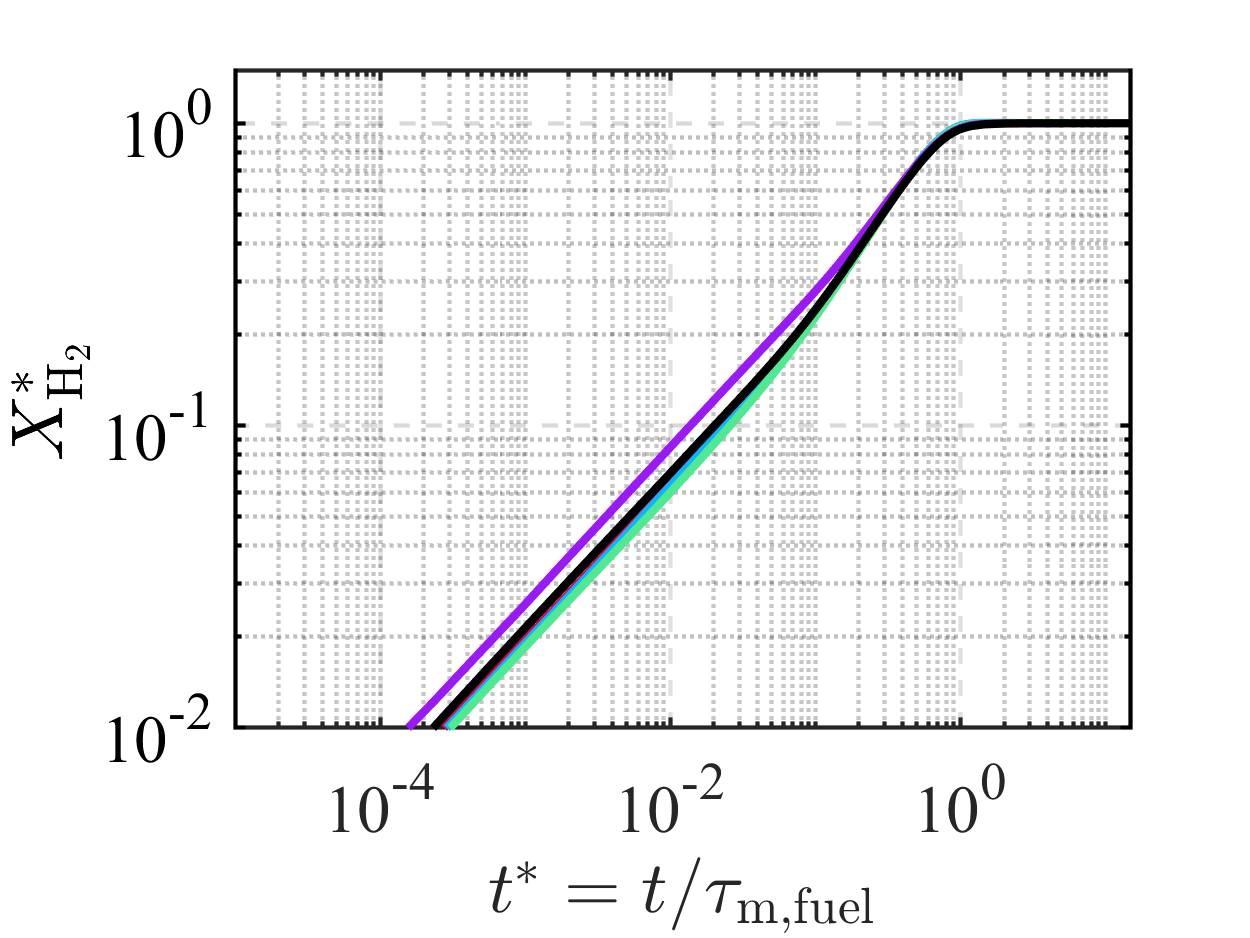}
        \caption{}
        \label{fig:SOFC_sub8}
    \end{subfigure}
    \hspace{-0.4cm}
    \begin{subfigure}[b]{0.32\textwidth}
        \includegraphics[width=\textwidth]{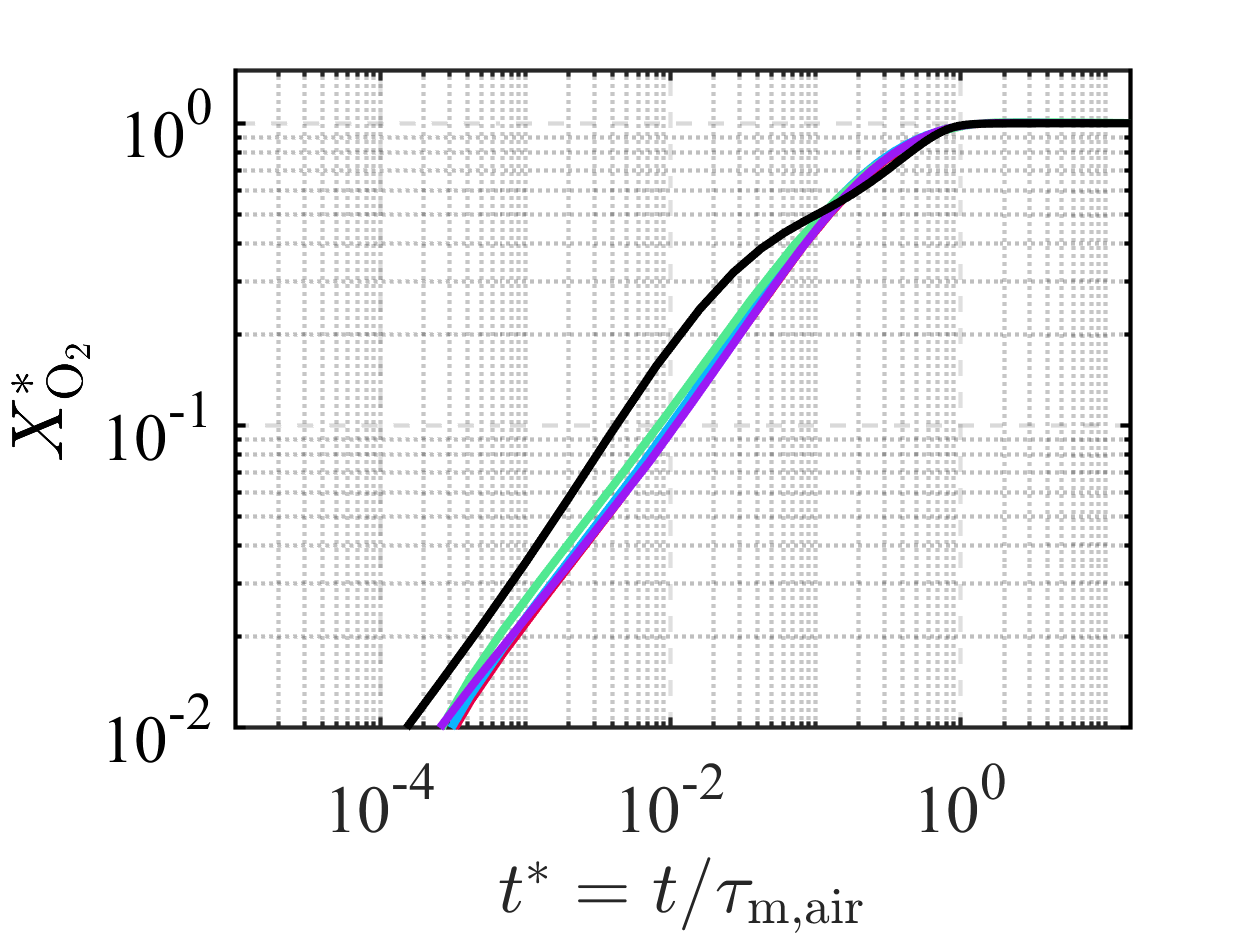}
        \caption{}
        \label{fig:SOFC_sub9}
    \end{subfigure}
    \caption{Transient responses of SOFC after step changes of current (mass source). SOFC with different dimensions, material properties, and operating conditions (`FC', `FC-T', `FC-I', and `FC-V') and the base case of SOEC are compared.}
    \label{fig:SOFC_transient}
\end{figure}

We extended our simulation test from SOECs to SOFCs, as shown in Fig.\,\ref{fig:SOFC_transient}. The operating conditions for SOECs and SOFCs differ significantly, as presented in Table\,\ref{Tab:ParaMass}. For example, SOFCs require large air flow rates to evacuate the heat generated by the electrochemical reactions \cite{BEALE2021100902}. Fig.\,\ref{fig:SOFC_sub1}--\subref{fig:SOFC_sub6} present the complex interplay between SOFC transients and various parameters, highlighting the stark contrast between SOEC and SOFC transients. By scaling the time with $\tau_{\rm m}$ in Fig.~\ref{fig:SOFC_sub7}, \ref{fig:SOFC_sub8}, and\ref{fig:SOFC_sub9}, we found that the relaxation times for H$_2$ and voltage are around $1 \tau_{\rm m}^{\rm fuel}$, and the relaxation time of O$_2$ is around $1 \tau_{\rm m}^{\rm air}$ after the step change of current. This finding further proves the effectiveness of $\tau_{\rm m}$ in characterizing mass-transfer transients for SOCs (including both SOECs and SOFCs). 

\subsubsection{Additional comments}
\begin{figure}[h]
    \centering
    \begin{subfigure}[b]{0.49\textwidth}
        \includegraphics[width=\textwidth]{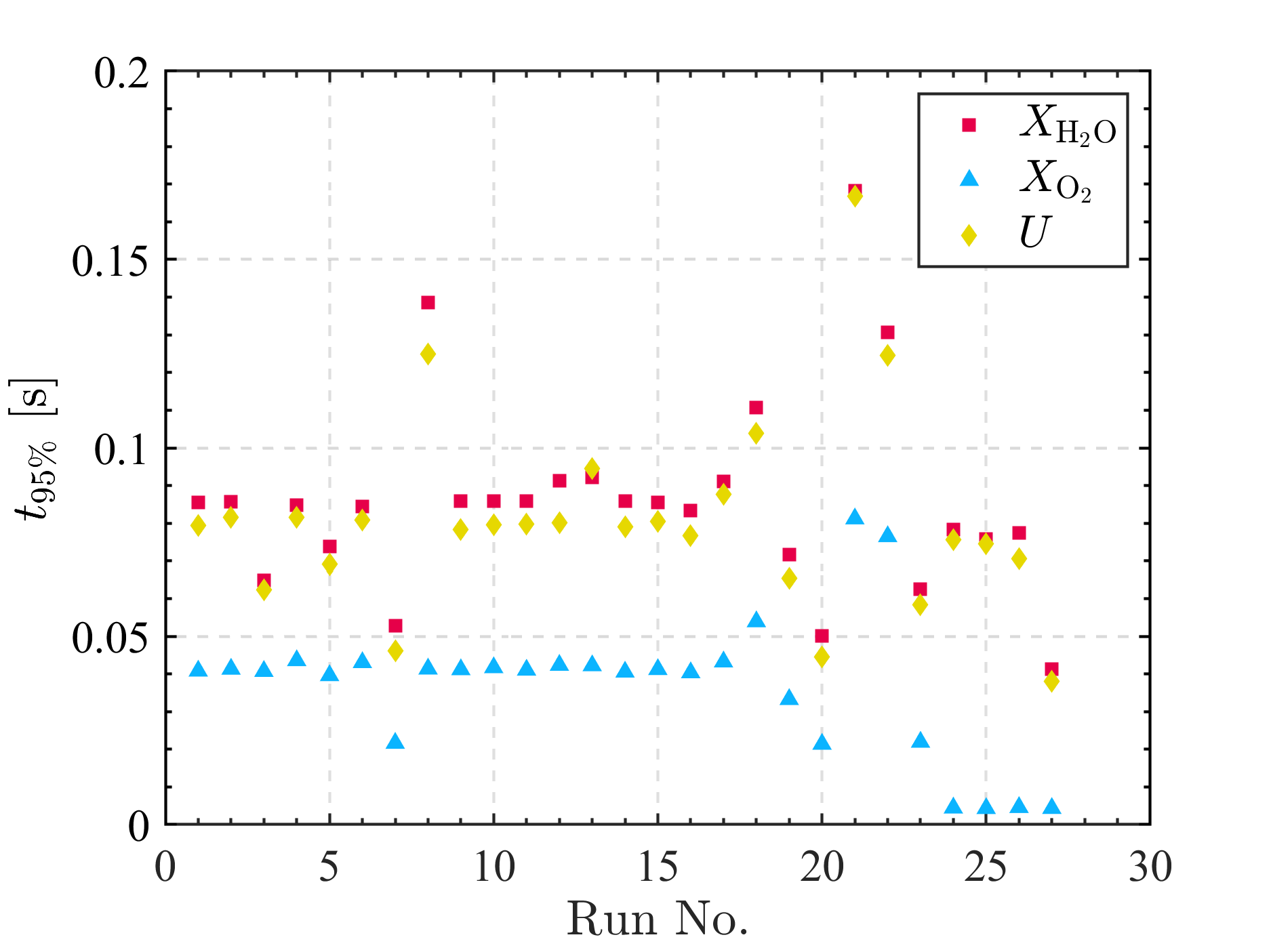}
        \caption{}
        \label{fig:Mass95_sub1}
    \end{subfigure}
    \hfill
    \begin{subfigure}[b]{0.49\textwidth}
        \includegraphics[width=\textwidth]{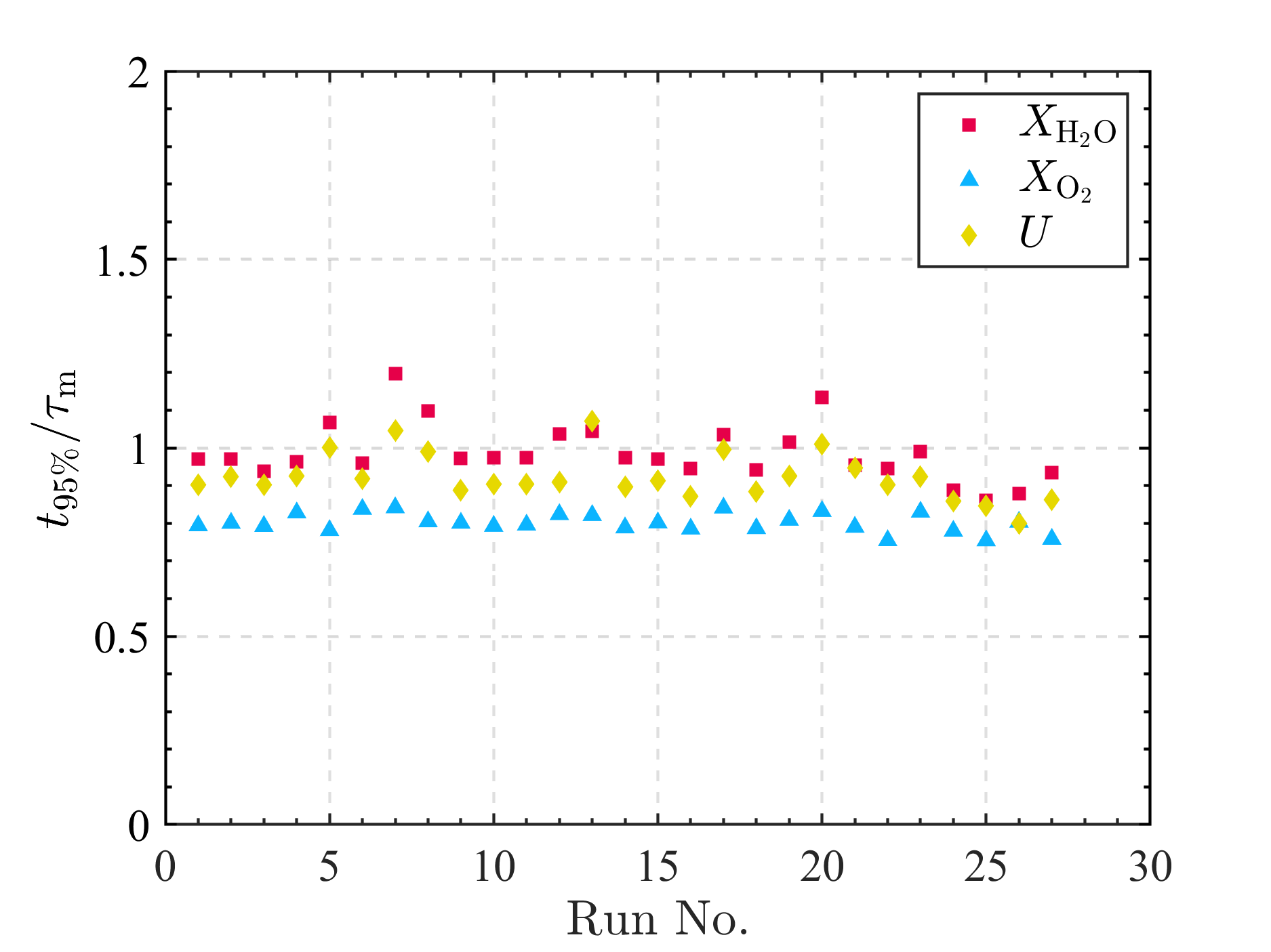}
        \caption{}
        \label{fig:Mass95_sub2}
    \end{subfigure}
    \caption{Time and dimensionless time required for voltage ($U$) and the mole fractions of H$_2$O and O$_2$ ($X_{\rm H_2O}$, $X_{\rm O_2}$) to reach 95\% steady state in different cases. The time to 95\% steady state is defined as the time at which $U^*$, $X_{\rm H_2O}^*$, or $X_{\rm O_2}^*$ equal to 0.95.}
    \label{fig:Mass95}
\end{figure}

Fig.\,\ref{fig:Mass95} provides a comparative overview of the relaxation times (defined as the time required to reach 95\% of steady state) for various cases. Voltage, fuel, and air take approximately $0.8 \tau_{\rm m}^{\rm fuel}$, $1 \tau_{\rm m}^{\rm fuel}$, and $0.9 \tau_{\rm m}^{\rm air}$ to reach 95\% of steady state. It is evident that the characteristic time $\tau_{\rm m}$ can provide a rough prediction for the relaxation times of SOEC and SOFC after step changes of current. 

\begin{figure}[h]
    \centering
    \begin{subfigure}[b]{0.49\textwidth}
        \includegraphics[width=\textwidth]{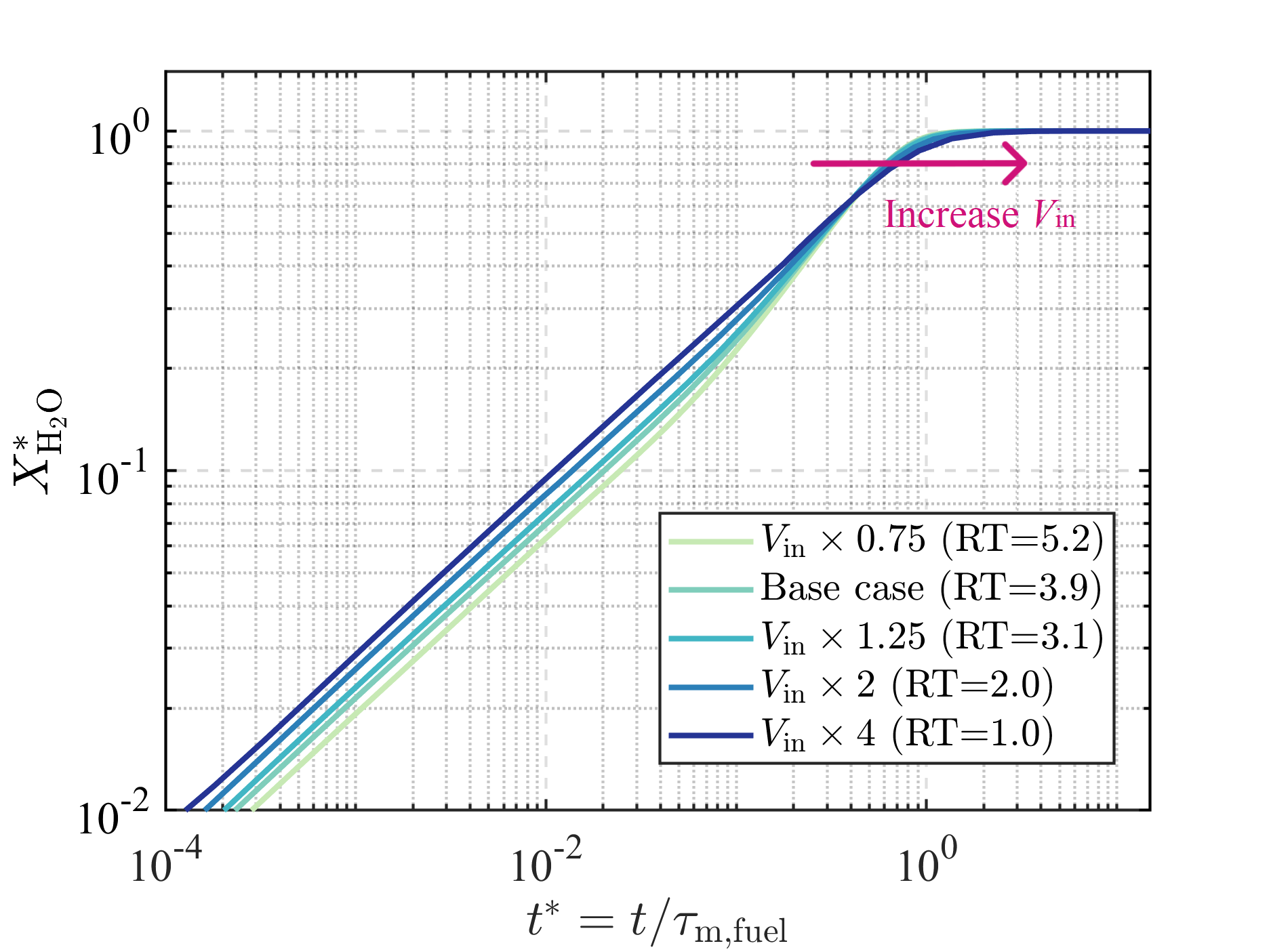}
        \caption{}
        \label{fig:comparedVel_sub1}
    \end{subfigure}
    \hfill
    \begin{subfigure}[b]{0.49\textwidth}
        \includegraphics[width=\textwidth]{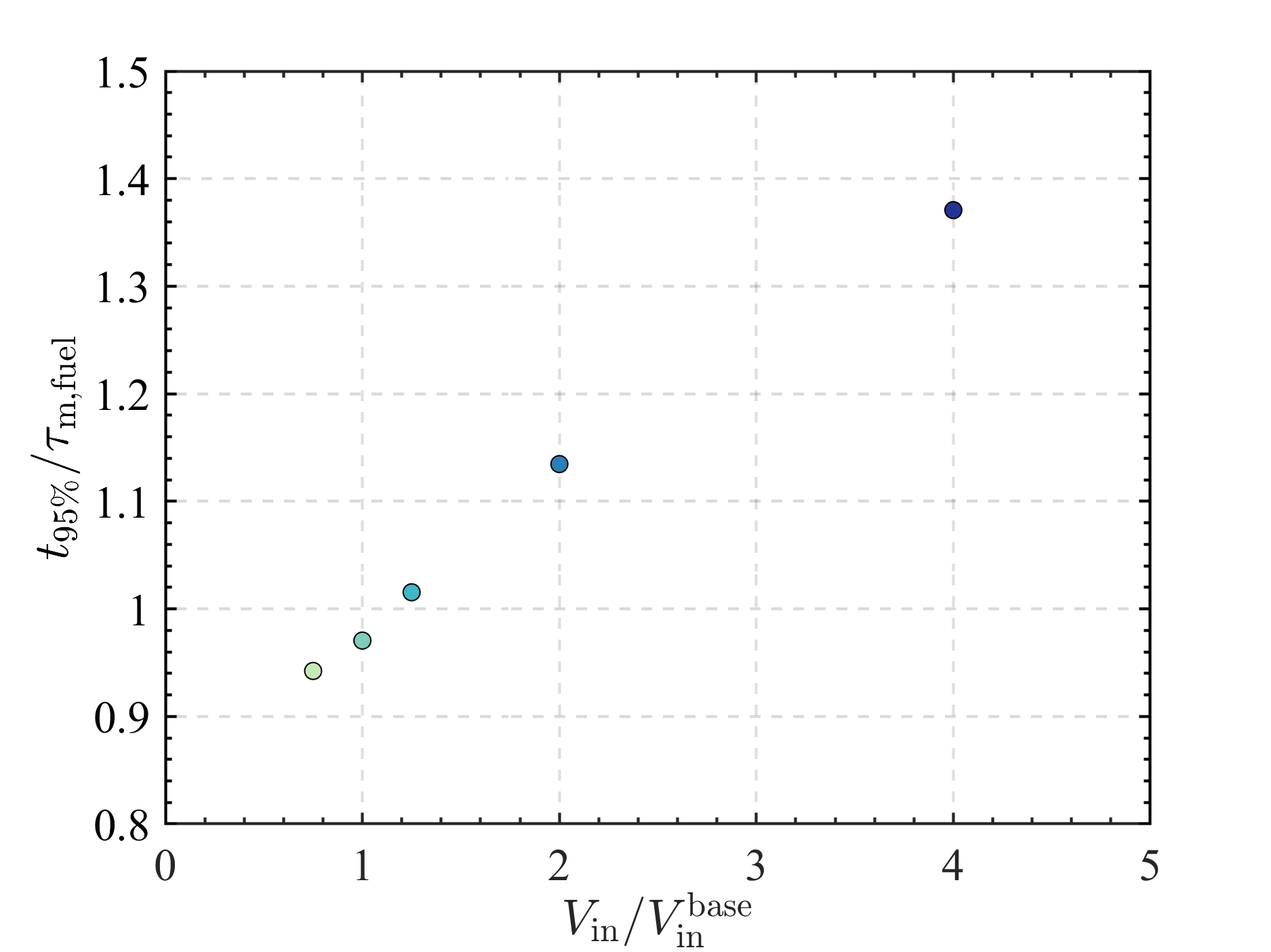}
        \caption{}
        \label{fig:comparedVel_sub2}
    \end{subfigure}
    \caption{Comparison on the transient responses of H$_2$O at different inlet velocities. `RT' indicates the ratio of mass-transfer characteristic time and the diffusion time constant of CDL in thickness direction, ${\rm RT} = \frac{\tau_{\rm m}}{\delta_{\rm{CDL}}^2 / D_{\rm{eff},0}}$.}
    \label{fig:comparedVel}
\end{figure}

However, it is important to note that the relationship between relaxation time and $\tau_{\rm m}$ or $\tau_{\rm t}$ is not a rigidly constant. For example, $X_{\rm H_2O}^*$ may take more than $1 \tau_{\rm m}^{\rm fuel}$ to reach 95\% of steady state if $\tau_{\rm m} \approx \tau_{\rm m,diff,\delta}$. As shown in Fig.\,\ref{fig:comparedVel}, when $V_{\rm in}$ is increased to four times the base case, the ratio of $\tau_{\rm m}$ and $\tau_{\rm m,diff,\delta}$ is reduced to around 1, and the relaxation time is increased to $1.4 \tau_{\rm m}^{\rm fuel}$ instead of remaining around $1 \tau_{\rm m}^{\rm fuel}$. This phenomenon can be explained by the fact that the overall mass-transfer time scale may result from the competition of $\tau_{\rm m}$ and $\tau_{\rm m,diff,\delta}$ in SOC and limited by the slowest sequence process. 

Considering the mass transfer in SOEC, the reactant H$_2$O must diffuse through the CDL to influence electrochemical reactions. However, the CDL serves as a supporting layer and is the thickest layer in the electrodes of the SOEC under investigation, thereby significantly hindering mass transfer. According to the non-dimensional analysis presented in Table\,\ref{Tab:Dimensional}, the diffusion time constant of CDL in the thickness direction $\tau_{\rm m,diff,\delta}$ is equal to $\delta_{\rm{CDL}}^2 / D_{\rm{eff},0}$. In the base case, $\tau_{\rm m}$ is 5.2 times greater than $\tau_{\rm m,diff,\delta}$, and thus limits overall mass transfer in SOEC. Whereas, increasing $V_{\rm in}$ reduces the value of $\tau_{\rm m}^{\rm fuel}$, as described in Eq.\,(\ref{eq:tau_m}). If $\tau_{\rm m}^{\rm fuel}$ is reduced to a value that approximates $\tau_{\rm m,diff,\delta}$, then the limiting effects of $\tau_{\rm m}^{\rm fuel}$ may no longer dominate the overall mass-transfer transients. Consequently, the relaxation time of $X_{\rm H_2O}^*$ will be significantly larger than $1 \tau_{\rm m}^{\rm fuel}$. In the actual operation of SOC, however, $\tau_{\rm m}<\tau_{\rm m,diff,\delta}$ is rare since it requires a very large velocity at fuel inlet, which typically results in low fuel utilization. Such low utilization rates are detrimental to efficient hydrogen production or power generation in SOCs.

\subsection{Characteristic time of heat transfer}\label{sec:charac_heat}
Temperature exerts a profound impact on SOC performance by affecting various properties such as the ionic conductance of the electrolyte, the diffusivity of gases, the activation energy of electrochemical reactions, and material properties \cite{njodzefon2013electrochemical}. Consequently, heat transfer within the SOC plays a crucial role in determining its transient performance. During the operation of SOEC, electrochemical reactions are endothermic below the thermal neutral voltage and exothermic vice versa. While under SOFC mode, electrochemical reactions are exothermic. The heat sources in SOEC and SOFC vary with voltage and current, introducing substantial complexity in determining the characteristic time of heat transfer. To reduce this complexity, we decoupled the heat sources from the electrochemical model to investigate the  characteristic time of heat transfer in SOEC and SOFC. A parametric test was conducted by manipulating the dimensions, thermal properties, and operating conditions of SOEC and SOFC, with the manipulated variables shown in Table\,\ref{Tab:ParaHeat}.

\begin{table}[h]
\centering
\small
\caption{Table of manipulated variables used in the parametric study of heat transfer.  A `–’ indicates that the parameter is consistent with the base case.}
\label{Tab:ParaHeat}
\begin{threeparttable}
\begin{tabular}{lllllllllllll}
\hline
\multirow{2}{*}{Name}       & \multirow{2}{*}{Run} & $k_{\rm DL}^{\rm fuel}$ & $c_{p,{\rm int}}$ & $H_{\rm ch}$     & $L_{\rm cell}$ & $T_{\rm in}$  & $V_{\rm in}^{\rm fuel}$ & $V_{\rm in}^{\rm air}$    & $X_{\rm in}^{\rm H_2O}$ & $X_{\rm in}^{\rm H_2}$ & $S_{{\rm h}, t=0}$      & $S_{{\rm h},t>0}$   \\ 
                          &     & [W/m.K]     & [J/kg.K]      & [m]          & [m]       & [K]     & [m/s]        & [m/s]               &                   &                  & [W/m$^3$]     & [W/m$^3$]     \\ \hline
Base   & 1   & 6     & 550               & 1.0E-03          & 0.10          & 1073.15    & 2.00             & 2.00                     & 0.70              & 0.30             & -2.83E+07         & -4.81E+07         \\
k1 & 2   & 0.6     & --      & --          & --            & --  & --             & --                     & --              & --             & --         & --         \\
k2 & 3   & 60     & --      & --          & --            & --  & --             & --                     & --              & --             & --         & --         \\
cp1 & 4   & --     & {225}      & --          & --            & --  & --             & --                     & --              & --             & --         & --         \\
cp2 & 5   & --     & {1100}      & --          & --            & --  & --             & --                     & --              & --             & --         & --         \\
H1     & 6   & --     & --               & {2.0E-03} & --           & --    & --             & --                    & --              & --             & --         & --         \\
L1    & 7   & --     & --               & --          & {0.05}   & --   & --             & --                    & --              & --             & --         & --         \\
L2    & 8   & --     & --               & --          & {0.20}   & --   & --             & --                    & --              & --             & --         & --         \\
T1     & 9   & --     & --               & --          & --          & {1073.15}  & --             & --              & --              & --             & --         & --         \\
T2     & 10   & --     & --               & --          & --          & {1123.15}  & --             & --              & --              & --             & --         & --         \\
V1     & 11   & --     & --               & --          & --           & --     & {4.00} & --                      & --              & --             & --         & --         \\
VXS          & 12   & --     & --               & --          & --         & --       & --    & {18.00}                & {0.10}     & {0.90}    & {4.83E+08} & {1.32E+08} \\ \hline
\end{tabular}
\end{threeparttable}
\end{table}

\begin{figure}[h]
    \centering
    \begin{subfigure}[b]{1\textwidth}
        \centering
        \includegraphics[height=5mm]{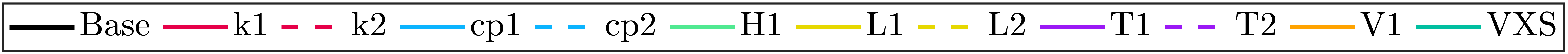}
    \end{subfigure}
    \\
    \begin{subfigure}[b]{0.32\textwidth}
        \includegraphics[width=\textwidth]{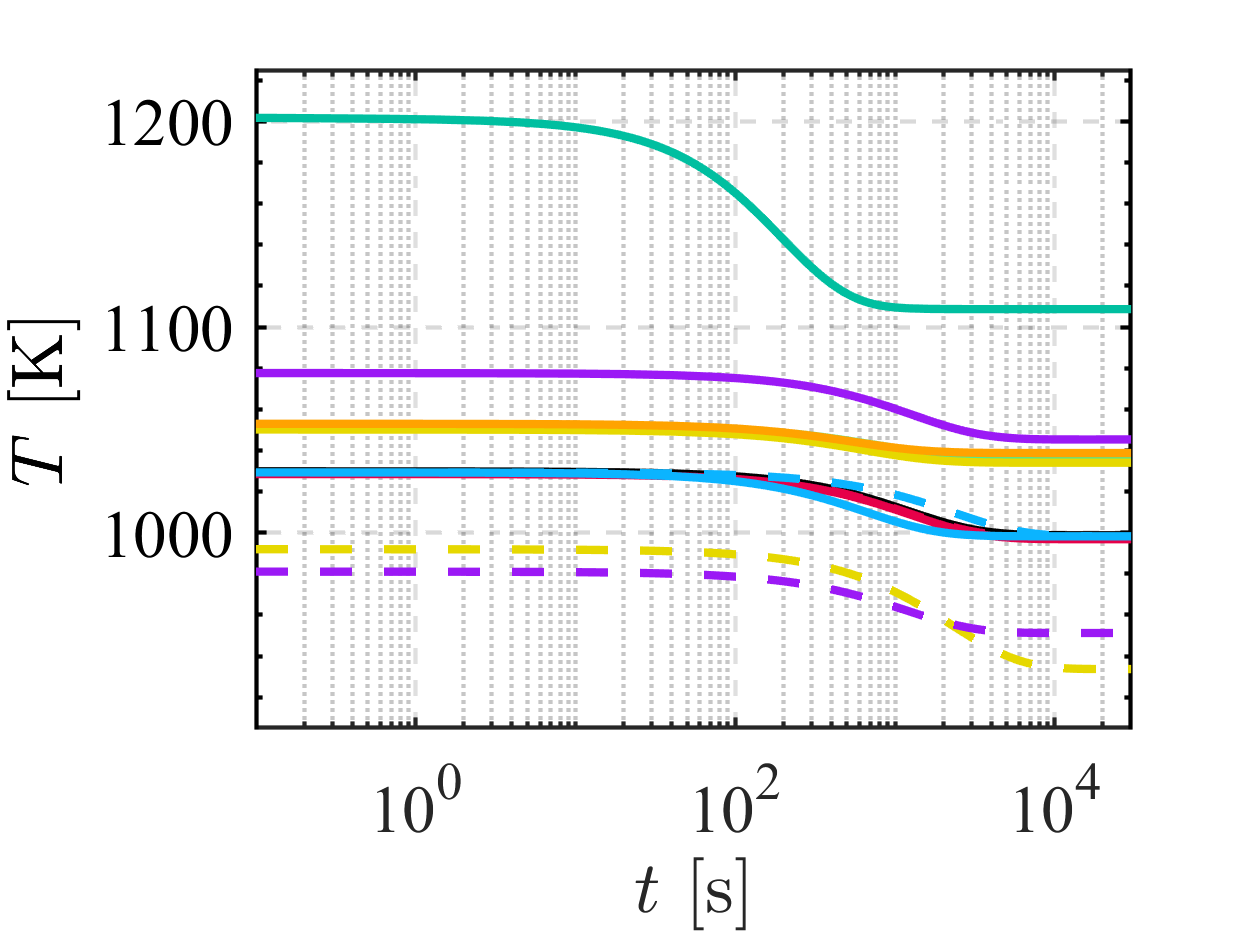}
        \caption{}
        \label{fig:thermal_sub1}
    \end{subfigure}
    \hspace{-0.4cm}
    \begin{subfigure}[b]{0.32\textwidth}
        \includegraphics[width=\textwidth]{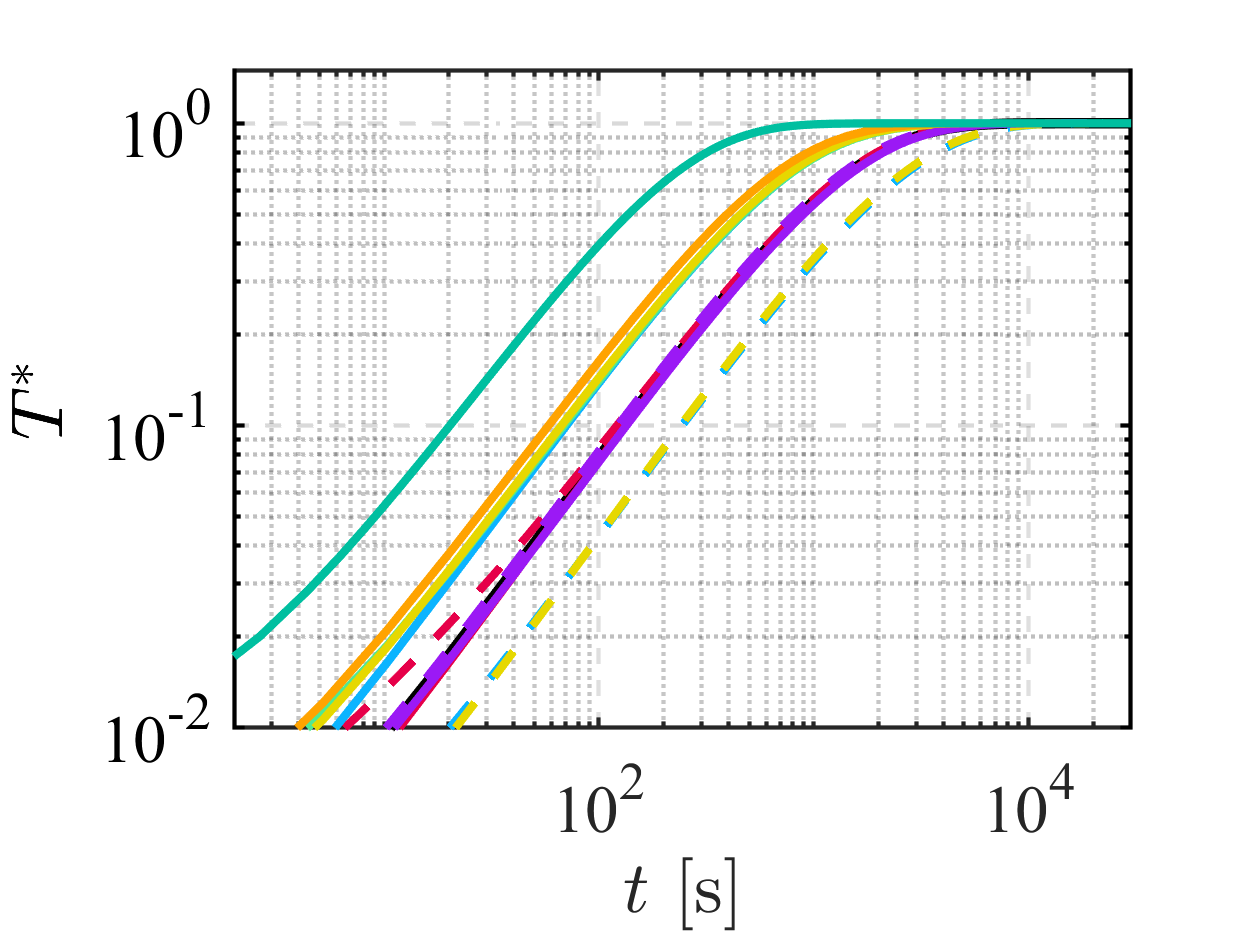}
        \caption{}
        \label{fig:thermal_sub2}
    \end{subfigure}
    \hspace{-0.4cm}
    \begin{subfigure}[b]{0.32\textwidth}
        \includegraphics[width=\textwidth]{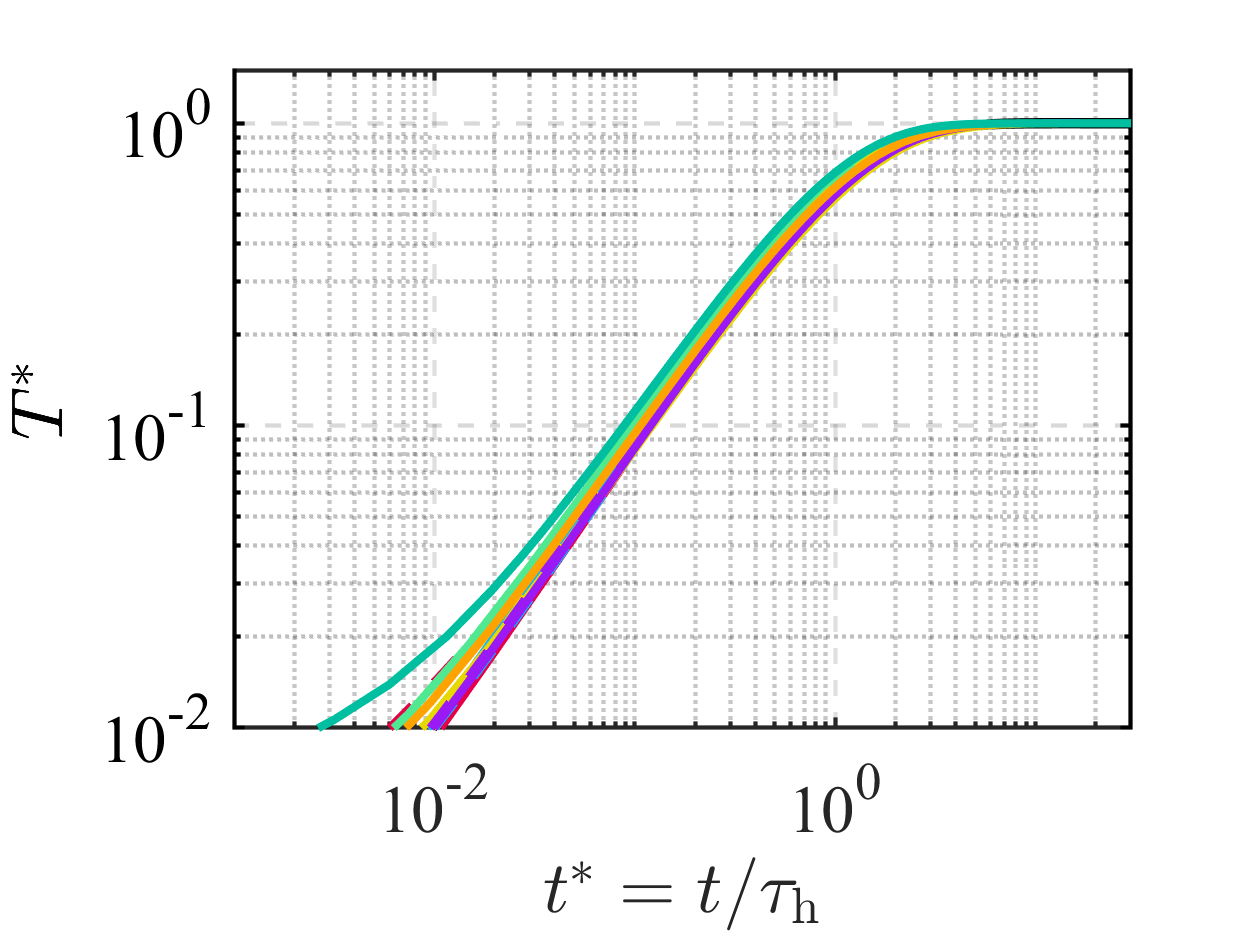}
        \caption{}
        \label{fig:thermal_sub3}
    \end{subfigure}
    \caption{Transient thermal responses after step changes of heat source. Temperatures of cells with different dimensions, material properties, and operating conditions are compared.}
    \label{fig:thermal_transient}
\end{figure}

Here, the thermal response of SOEC or SOFC following step changes of heat sources is of interest. Each simulation case is initialized with its steady state and a constant heat source $S_{{\rm h}, t=0}$ in AFL and CFL, considering that the majority of heat source or heat sink is induced by the electrochemical reactions in FL. The heat source is then changed to $S_{\rm h,t>0}$ in a small time step ($<0.01\,$s) and held until the end of simulation. During the simulation, average temperature of FL in the SOEC or SOFC are monitored. The simulation results are presented in Fig.\,\ref{fig:thermal_sub1} and \ref{fig:thermal_sub2}. The dimensionless temperature $T^*$, defined in Table\,\ref{Tab:Dimensional}, represents the extent to which steady state has been achieved. $T^*=1$ indicates the temperature has reached the steady state. With different  dimensions, thermal properties, and operating conditions, the relaxation time of SOC after a step change in heat source varies widely from $10^{2}$\,s to $10^{3}$\,s. 

To elucidate the relationship between thermal transients and various parameters, we proposed the characteristic time of heat transfer within SOCs, denoted as $\tau_{\rm h}$. Upon scaling the time with $\tau_{\rm h}$, the curves of $T^*$--$t^*$ converge to a similar trend in Fig.\,\ref{fig:thermal_sub3}, showing a comparable relaxation time of around $2\tau_{\rm h}$. These results demonstrate the effectiveness of $\tau_{\rm h}$ in representing the overall heat-transfer time scale of SOEC and SOFC.  

The expression of $\tau_{\rm h}$, analogous to the characteristic time of mass transfer Eq.\,(\ref{eq:tau_m}), is expressed as Eq.(\,\ref{eq:tau_th}).
%\begin{equation} \label{eq:tau_th}
\begin{align}\label{eq:tau_th}
\tau_{\text h}&=\frac{\text { Total enthalpy }}{\text { Total heat transfer rate }} \frac{[\rm{J}]}{[\rm{W}]} \nonumber \\
&\approx \frac{\left(\left(m c_p\right)_{\rm{int}}^{\text {solid }}+\left(m c_p\right)_{\rm{E}}^{\rm{solid}}+\left(m c_p\right)_{\rm{AFL}}^{\rm{eff}}+\left(m c_p\right)_{\rm{ADL}}^{\rm{eff}}+\left(m c_p\right)_{\rm{CDL}}^{\rm{eff}}+\left(m c_p\right)_{\rm{CFL}}^{\rm{eff}}+\left(m c_p\right)_{\rm{ch}}^{\rm{fluid}}\right)\times T_{\rm in}}{\dot{\mathcal{H}}_{\rm{in}}^{\rm{fuel}}+\dot{\mathcal{H}}_{\rm{in}}^{\rm{air}}}
\end{align}
%\end{equation}
where, $m$ denotes the mass, $\dot{V}$ denotes the volumetric flow rate, the superscript `eff' indicates the properties of porous media that are calculated by volumetrically averaging the fluid and solid properties, and $\dot{\mathcal{H}}$ denotes the enthalpy flow rate of fluid. The physical meaning of $\tau_{\rm h}$ is the time required to `fill' the total enthalpy of the system via the heat transfer rate through the system. Given the boundary conditions of the system under investigation (as per Table\,\ref{Tab:BC}), there are only two inlets and two outlets that exchange heat with the environment, while the other boundaries are adiabatic. Thus, the total heat transfer rate of the system can be approximated by summing the enthalpy flow rates of fluids at the two inlets. This approximation may induce minor errors in calculating $\tau_{\rm h}$ since $\dot{\mathcal{H}}$ differs throughout the system owing to the internal heat source/sink. Nonetheless, this approximation remains reasonable for the cases under study as changes in $\dot{\mathcal{H}}$ inside the system can be considered negligible compared to $\dot{\mathcal{H}}_{\rm in}$ at inlets.

By utilizing the expression for $\tau_{\rm h}$ in Eq.~(\ref{eq:tau_th}), it is possible to determine the relative importance of various parameters on the thermal transients of SOCs. As shown in Fig\,\ref{fig:composition_heat}, the interconnect possesses the largest heat capacity among all components within the studied SOC, and thus exerts a dominant influence on the value of $\tau_{\rm h}$. Besides, fluid flow rates, as the numerator of Eq.~(\ref{eq:tau_th}), are a dominating factor in the thermal response time, which is in alignment with the findings of Liu et al. \cite{LIU2022115318}'s study on thermal control.

\subsection{Validation of the proposed characteristic time}
To validate the proposed characteristic times of heat and mass transfer, we applied Eqs.\,(\ref{eq:tau_m}) and (\ref{eq:tau_th}) to calculate the $\tau_{\rm m}$ and $\tau_{\rm h}$ of SOCs and a PEMFC that are reported in literature. The calculated $\tau_{\rm m}$ and $\tau_{\rm h}$ were then compared with the reported relaxation times. As shown in Table\,\ref{Tab:ValidationTime}, the calculated $\tau_{\rm m}$ and $\tau_{\rm h}$ and the reported relaxation times are of the same order of magnitude despite differences in application scenarios, such as planar/tubular SOCs, co-electrolysis, and even PEMFC. This validation not only further confirms the generalizability of the proposed characteristic times in representing the dynamic gaseous and thermal responses of various SOCs, but also opens up the possibility of characterizing the dynamic processes of other electrochemical cells such as PEMFCs. 

\begin{threeparttable}[h]
\caption{Comparison between the calculated characteristic times with the reported relaxation times of variables in responses to rapid voltage/current changes. }
\label{Tab:ValidationTime}
\begin{tabular}{llll}
\hline
Cell type                                                                        & Response variables             & Reported relaxation time [s]                                                                                     & Calculated $\tau_{\rm m}$, $\tau_{\rm h}$ [s]                                        \\ \hline
3-D Planar SOFC \cite{BAE2018405}                                                                  & Current           &  $\approx 0.6$ $^{\rm a}$                                                                                      & $\tau_{\rm m}^{\rm fuel}$= 0.46                                                      \\
\begin{tabular}[c]{@{}l@{}}2-D tubular SOEC  \cite{LUO2015637}\\ (co-electrolysis)\end{tabular}     & \begin{tabular}[c]{@{}l@{}} Channel H$_2$O concentration \\ Temperature \end{tabular}  & \begin{tabular}[c]{@{}l@{}} $0.255$ -- $1$  $^{\rm b}$\\ $515$ -- $2000$  $^{\rm b}$\end{tabular} & \begin{tabular}[c]{@{}l@{}}$\tau_{\rm m}^{\rm fuel}$= 0.42\\ $\tau_{\rm h}= 2076$\end{tabular} \\
3-D Planar SOFC \cite{NERAT2017728}                                                                 & \begin{tabular}[c]{@{}l@{}} Fuel utilization  \\ Air utilization \end{tabular}           & \begin{tabular}[c]{@{}l@{}} $\approx 0.7$ $^{\rm a}$\\ $\approx 0.05$ $^{\rm a}$\end{tabular}   & \begin{tabular}[c]{@{}l@{}}$\tau_{\rm m}^{\rm fuel}$= 0.49\\ $\tau_{\rm m}^{\rm air} = 0.04$ \end{tabular}  \\
\begin{tabular}[c]{@{}l@{}}Quasi 2-D SOEC stack \cite{BANERJEE2018996} \\ (co-electrolysis)\end{tabular} & \begin{tabular}[c]{@{}l@{}}Current \\ Stack temperature \end{tabular}    & \begin{tabular}[c]{@{}l@{}} $<1$  $^{\rm b}$\\ $>900$  $^{\rm b}$\end{tabular}          & \begin{tabular}[c]{@{}l@{}}$\tau_{\rm m}^{\rm fuel}= 0.23$\\ $\tau_{\rm h}=3100$ \end{tabular} \\
\begin{tabular}[c]{@{}l@{}}Flat-tube SOC \cite{WU2023143275} \\ (CO$_2$ electrolysis)\end{tabular} & Electrical impedance    &  $\approx 1.0$ $^{\rm c}$         & $\tau_{\rm m}^{\rm fuel}= 0.97$ \\
\begin{tabular}[c]{@{}l@{}}Planar PEMFC \cite{CHO2008118}\\ (experiment)\end{tabular}              & Current           &  $\approx 1.0$ $^{\rm a}$                                                                                        & $\tau_{\rm m}^{\rm fuel}= 0.41$                                                      \\ 
\hline
\end{tabular}
\begin{tablenotes}
\footnotesize
\item[a] The relaxation time is interpreted from figures.
\item[b] The relaxation time is interpreted from text and figures.
\item[c] The relaxation time is derived from Distribution of Relaxation Time (DRT) figures, with detailed parameters for calculating $\tau_{\rm m}^{\rm fuel}$ provided by Guan, the corresponding author in \cite{WU2023143275}. We would like to thank Guan for sharing the data.
\end{tablenotes}
\end{threeparttable}

\section{Applications of the characteristic time} \label{Sec:application}
The characteristic times of heat and mass transfer, $\tau_{\rm h}$ and $\tau_{\rm m}$, derived in Section\,\ref{Sec:parametric} are useful not only in the prediction of relaxation time, but also in guiding the design and control of SOEC and SOFC with minimal computational effort. In this section, we present two examples that demonstrate the practical applications of the proposed characteristic times, emphasizing their simplicity and effectiveness.

\subsection{Adjustment of design}\label{sec:optimization}
\begin{figure}[h]
    \centering
    \begin{subfigure}[b]{0.49\textwidth}
        \includegraphics[width=0.8\textwidth]{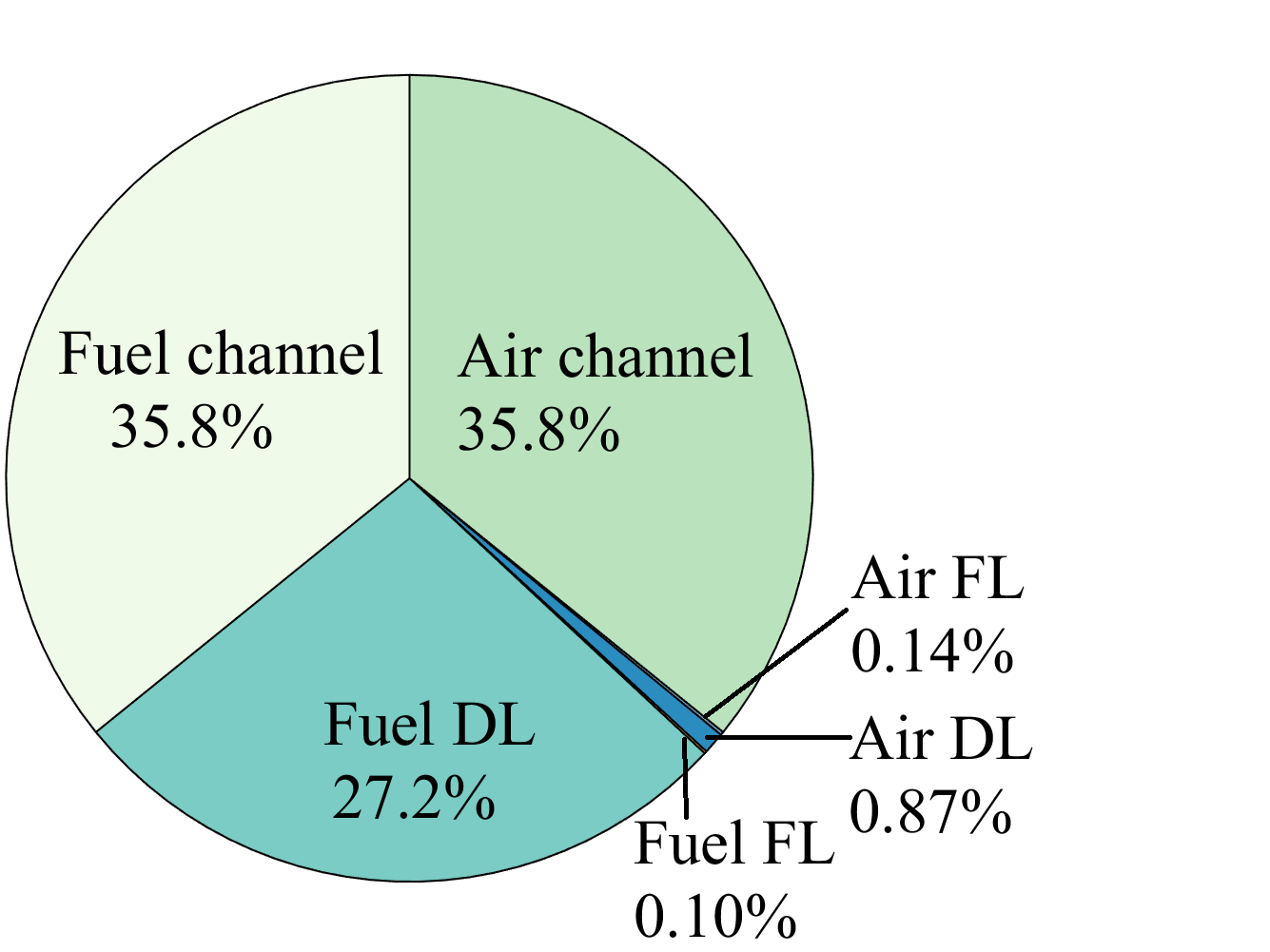}
        \caption{Compositions of total void volume.}
        \label{fig:composition_void}
    \end{subfigure}
    \hfill
    \begin{subfigure}[b]{0.49\textwidth}
        \includegraphics[width=0.8\textwidth]{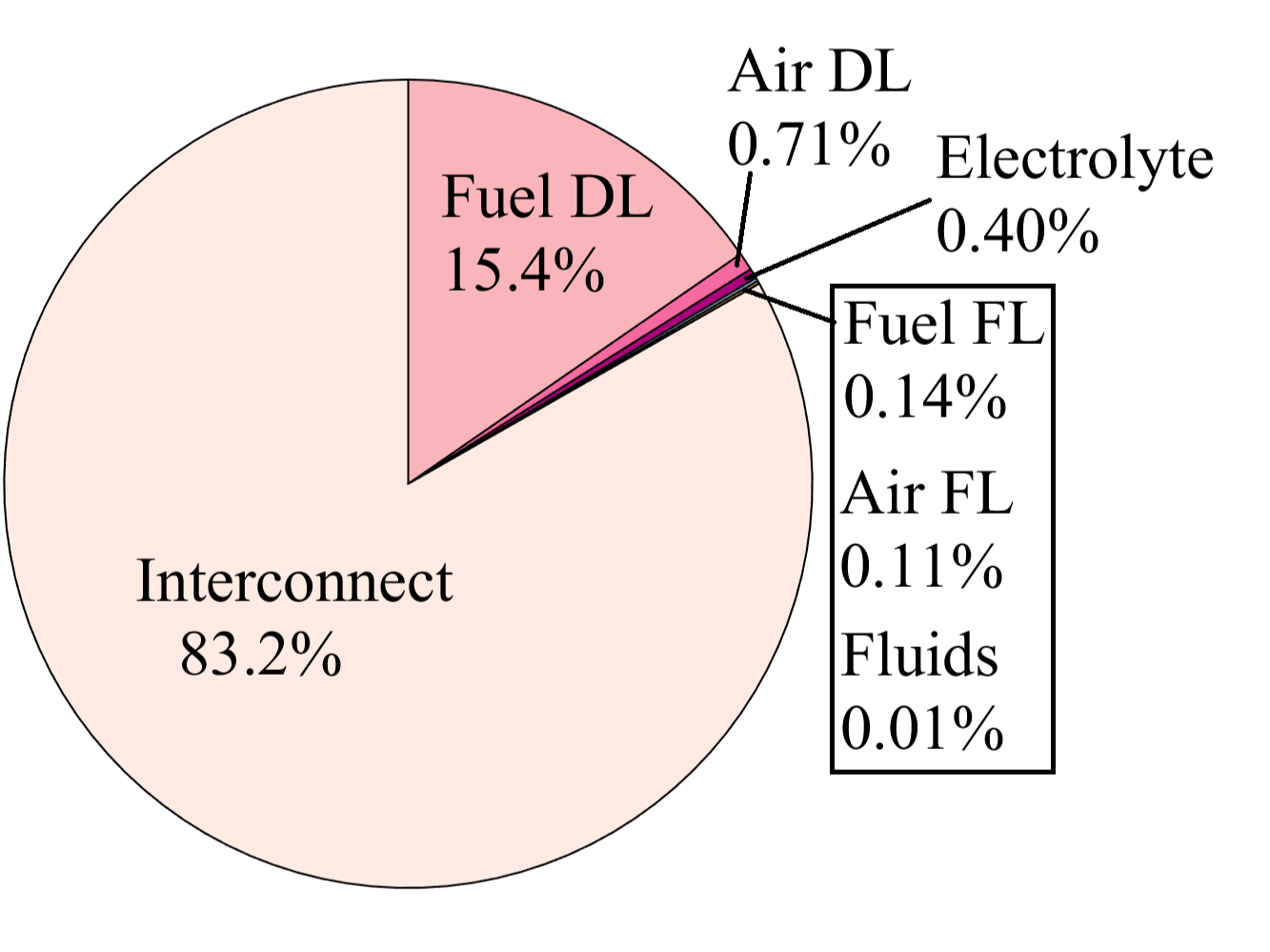}
        \caption{Compositions of total heat capacity.}
        \label{fig:composition_heat}
    \end{subfigure}
    \caption{Compositions of void volume and heat capacity in the base-case SOC.}
    \label{fig:composition}
\end{figure}

The design of SOC involes a multitude of parameters. By utilizing the simple expressions of $\tau_{\rm h}$ and $\tau_{\rm m}$, it is possible to identify the key parameters influencing SOC transients and subsequently design an SOC with the desired transient characteritics. Taking the base case of SOEC (Table\,\ref{Tab:ParaMass}) as an example, $H_{\rm ch}$ is crucial in determining $\tau_{\rm m}$ since the fluid channel contributes the largest void volume within the SOC, as shown in Fig.\,\ref{fig:composition_void}. Consequently, in the adjusted SOEC, $H_{\rm ch}$ was halved and $V_{\rm in}^{\rm fuel}$ was doubled to reduce the void volume while maintaining the volumetric flow rate and fuel utilization. This resulted in a smaller value for $\tau_{\rm m}^{\rm fuel}$ and a faster gaseous response for the adjusted SOEC. In addition, Fig.\,\ref{fig:composition_heat} shows that the interconnect has the largest heat capacity within the SOC in the base case. According to Eq.\,(\ref{eq:tau_th}), the total enthalpy and enthalpy flow rate are heavily dependent on $c_{p,{\rm int}}$ and $V_{\rm in}^{\rm air}$, respectively. Therefore, in the adjusted SOEC, $V_{\rm in}^{\rm air}$ was increased tenfold and $c_{p,{\rm int}}$ was halved to reduce $\tau_{\rm h}$ and accelerate the thermal response. 

\begin{figure}[h]
    \centering
    \begin{subfigure}[b]{1\textwidth}
        \includegraphics[width=\textwidth]{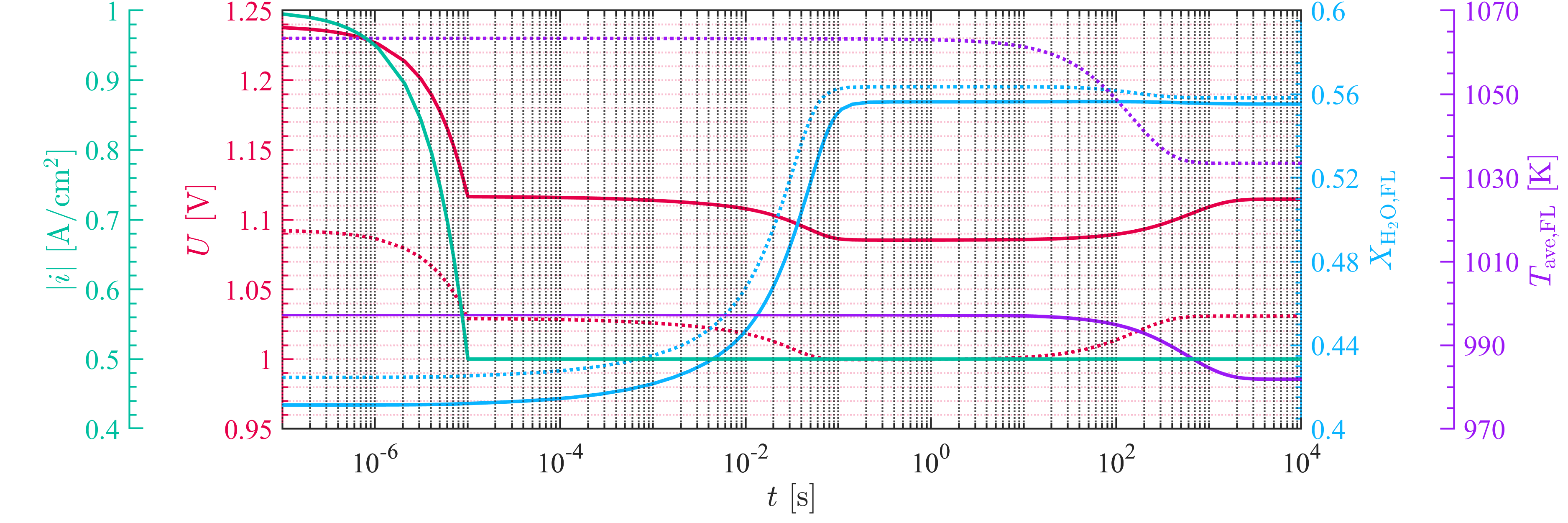}
        \caption{Responses in logarithmic timeline.}
        \label{fig:optimization_log}
    \end{subfigure}
    \\
    \begin{subfigure}[b]{0.49\textwidth}
        \includegraphics[width=\textwidth]{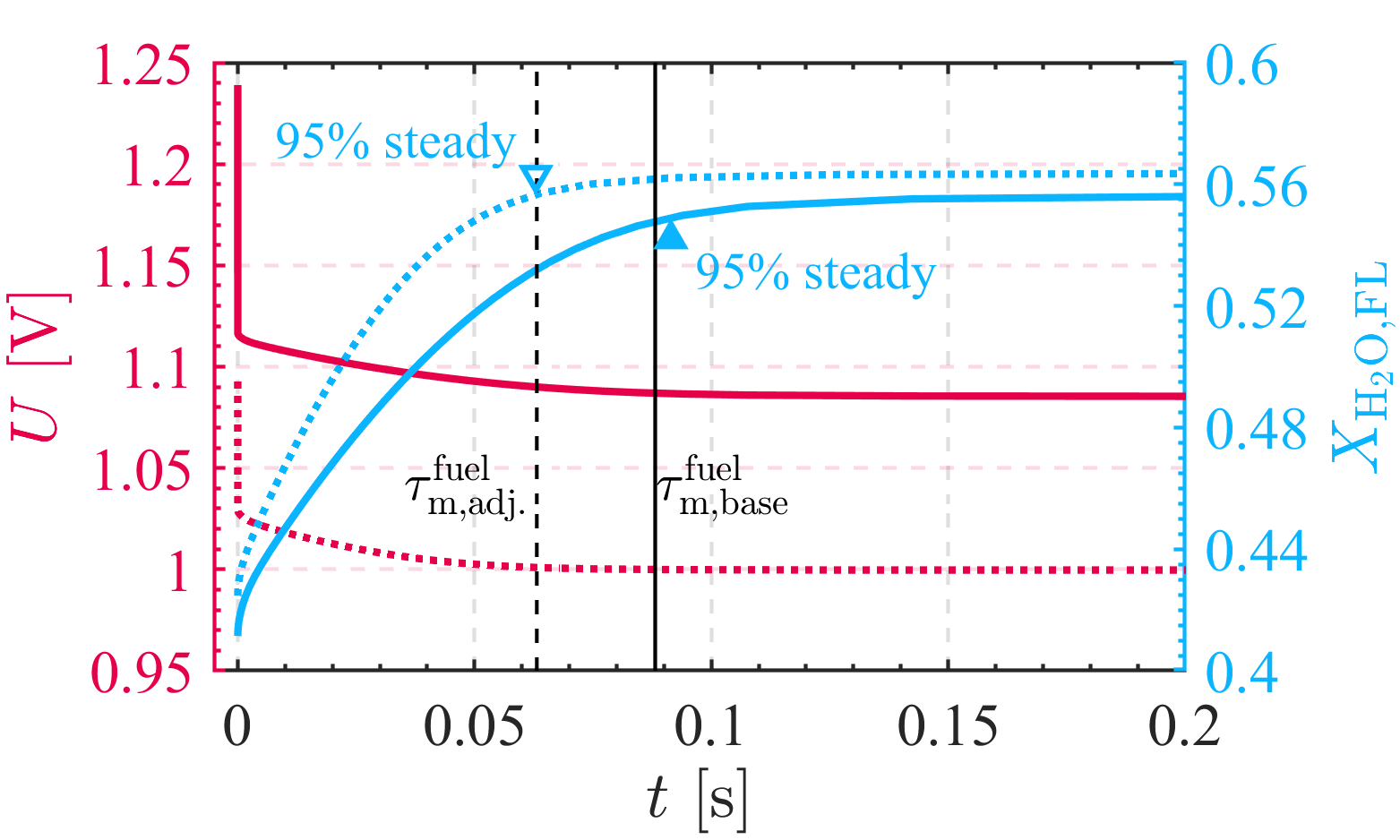}
        \caption{Responses in the linear timeline from 0\,s to 0.1\,s.}
        \label{fig:optimization_linear1}
    \end{subfigure}
    \hfill
    \begin{subfigure}[b]{0.49\textwidth}
        \includegraphics[width=\textwidth]{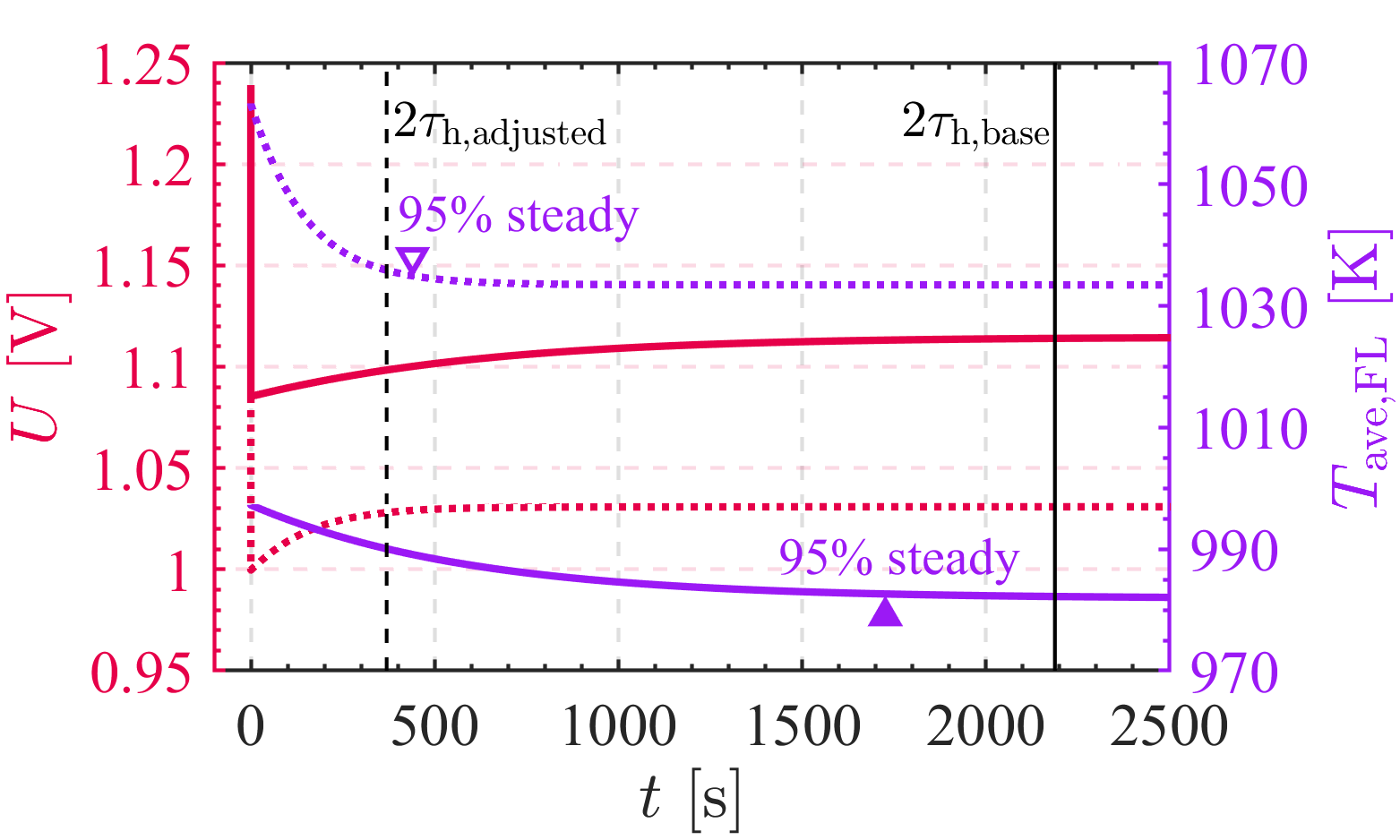}
        \caption{Responses in the linear timeline from 0\,s to 2500\,s.}
        \label{fig:optimization_linear2}
    \end{subfigure}
    \caption{Comparison on the transient responses of two SOEC after the step change of current. Solid line represents the base case of SOEC, and the dashed line represents the adjusted SOEC.}
    \label{fig:optimization}
\end{figure}

Fig.\,\ref{fig:optimization} compares the responses between the base and adjusted cases. In response to the step change of current magnitude $|i|$ from 1\,A to 0.5\,A, both cases exhibit delayed responses of $X_{\rm H_2O}$ and $T$ due to heat and mass transfer lags \cite{LIANG2023116759}. The adjusted SOEC exhibits significantly faster electrical, gaseous, and thermal responses. The relaxation time of $X_{\rm H_2O}$ decreased from 0.09\,s to 0.06\,s (Fig.\,\ref{fig:optimization_linear1}), and the relaxation time of $T$ decreased from 1700\,s to 440\,s (Fig.\,\ref{fig:optimization_linear2}). Notably, the relaxation time of $X_{\rm H_2O}$ is close to the value of $\tau_{\rm m}^{\rm fuel}$ in both cases, indicating the potential for characteristic times to serve as a precise and quantifiable indicator in the design of SOC. In Fig.\,\ref{fig:optimization_linear2}, the relaxation times of $T$ differ slightly from $2\tau_{\rm h}$ in both case. This fact implies that $\tau_{\rm h}$ can only provide a rough estimate of the thermal relaxation time following a step change of current. This is due to the complexity introduced by the varying heat source of SOEC in response to dynamic electrical conditions.

\subsection{Control strategy}
\begin{figure}[h]
    \centering
    \begin{subfigure}[b]{1\textwidth}
        \centering
        \includegraphics[height=5mm]{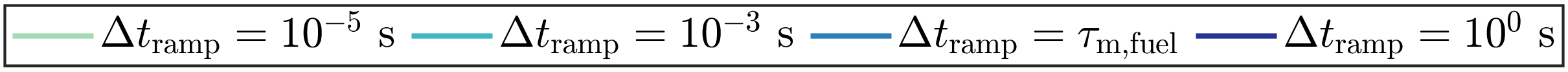}
    \end{subfigure}
    \\
    \begin{subfigure}[b]{0.32\textwidth}
        \includegraphics[width=\textwidth]{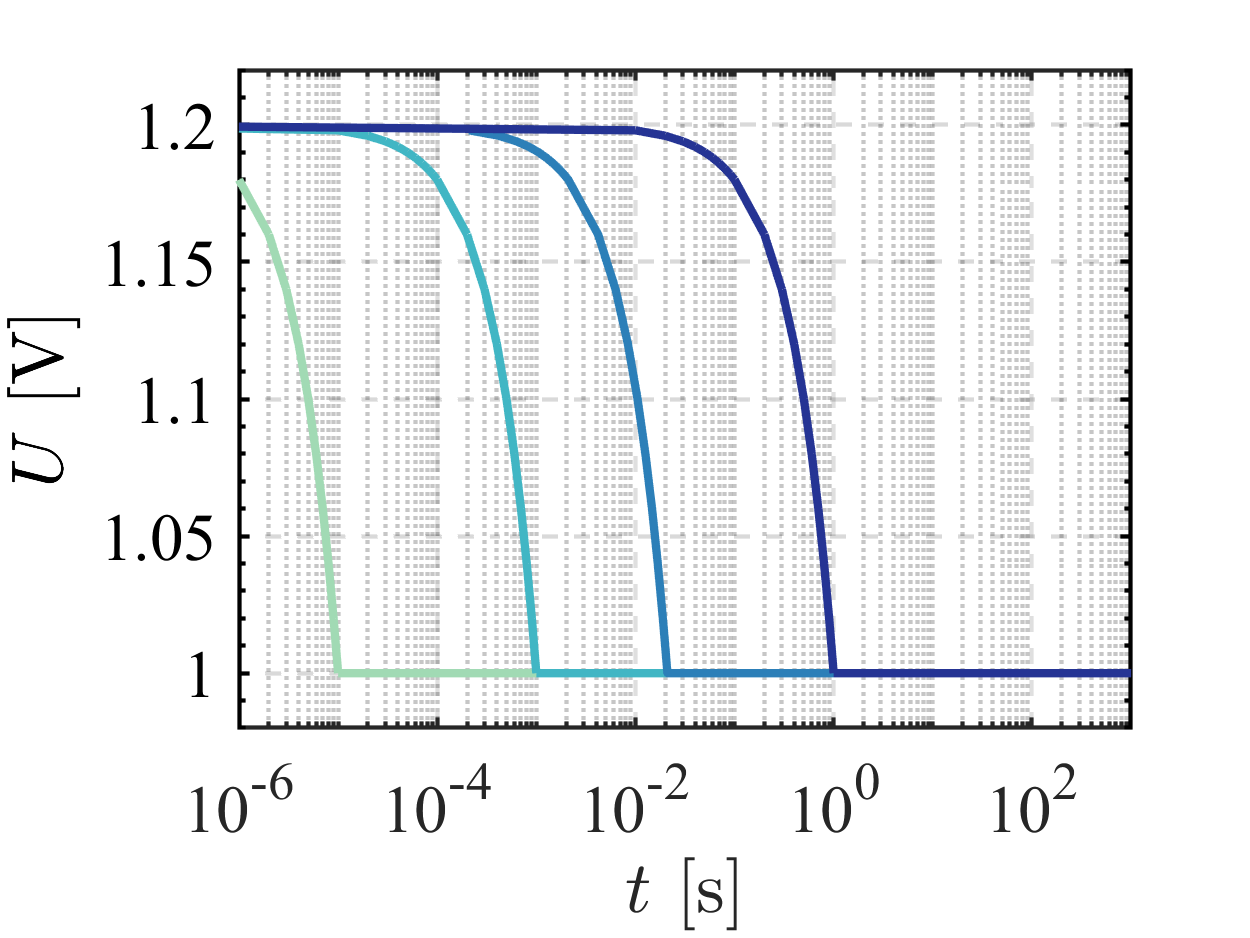}
        \caption{}
        \label{fig:control_sub1}
    \end{subfigure}
    \hfill
    \begin{subfigure}[b]{0.32\textwidth}
        \includegraphics[width=\textwidth]{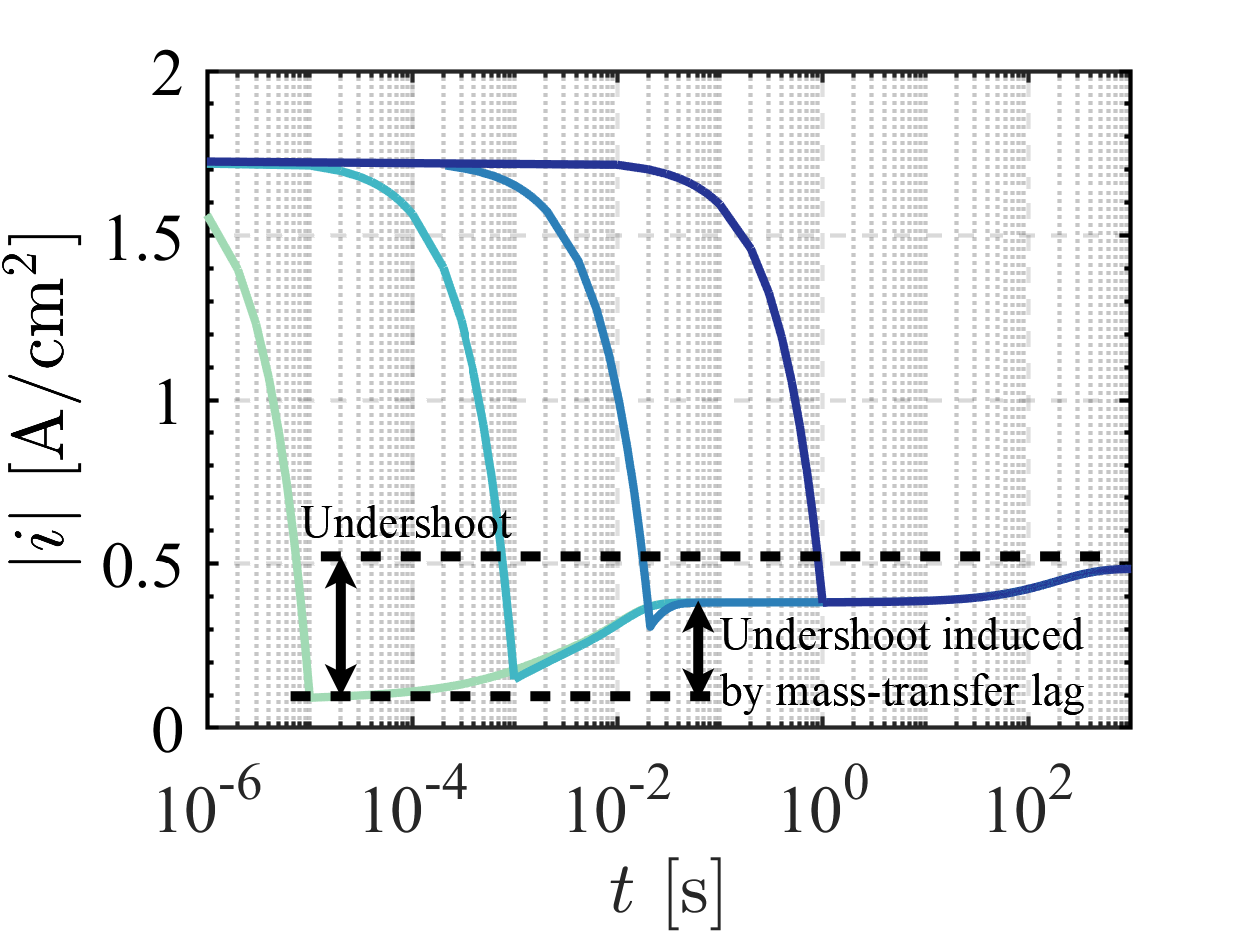}
        \caption{}
        \label{fig:control_sub2}
    \end{subfigure}
    \hfill
    \begin{subfigure}[b]{0.32\textwidth}
        \includegraphics[width=\textwidth]{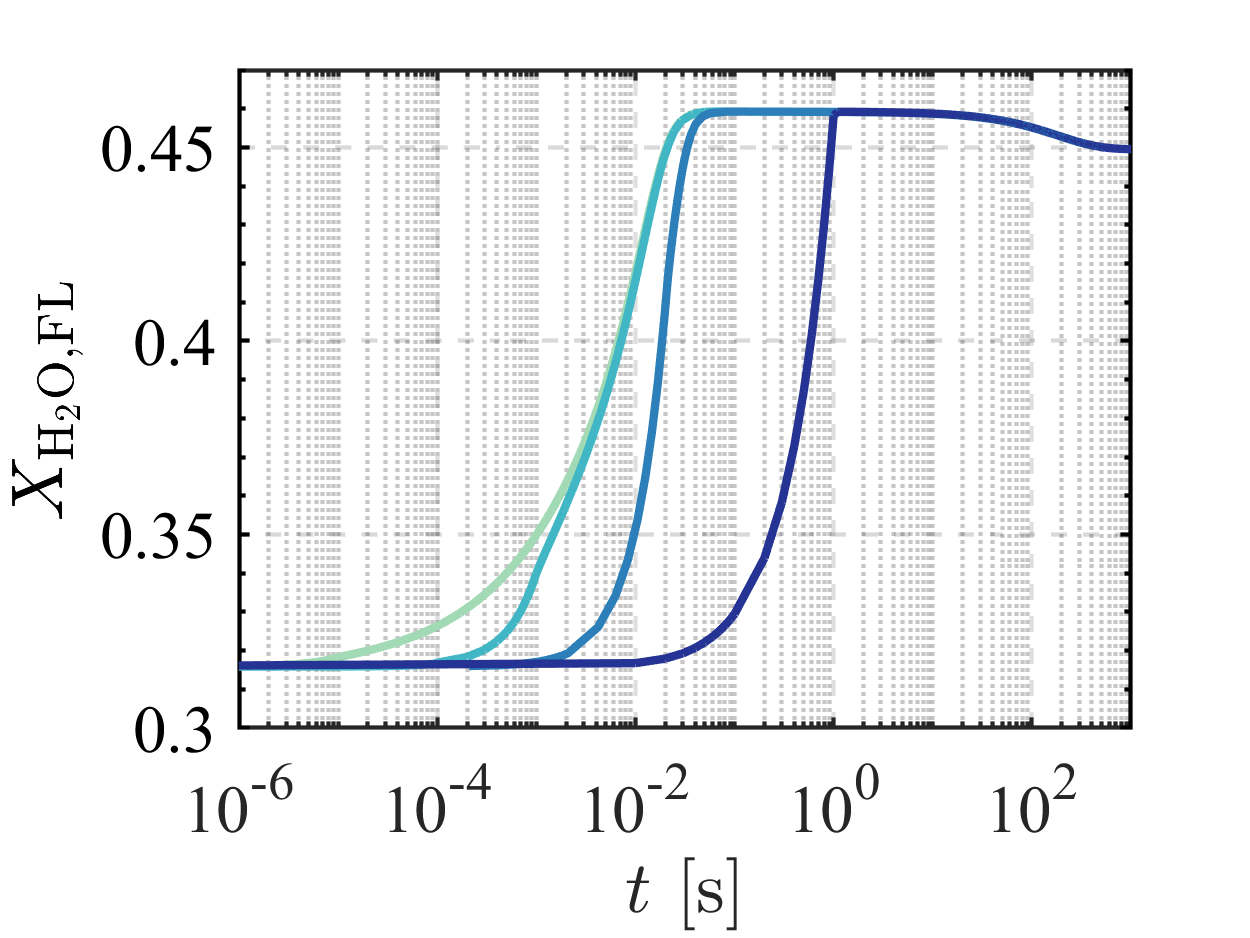}
        \caption{}
        \label{fig:control_sub3}
    \end{subfigure}
    \caption{Current responses of SOEC to current ramps with different ramp rates \cite{LIANG2023116759}.}
    \label{fig:control}
\end{figure}

In our previous study \cite{LIANG2023116759}, we found that heat and mass transfer lags can cause current undershoot after rapid voltage ramps on SOEC. We suggested reducing the voltage ramp to a certain ramp time to mitigate current undershoot induced by mass-transfer lag, as shown in Fig.\,\ref{fig:control}. However, the suggested ramp time would vary for different SOCs. It could not be determined without conducting experiments or simulations. This limitation impairs the generalizability of the control strategy for SOC. Our current study provides a convenient method for estimating the ramp time required to alleviate undershoot using Eq.~(\ref{eq:tau_m}). The ramp time is recommended to be larger than $1\tau_{\rm m}^{\rm fuel}$, equivalent to the relaxation time of $X_{\rm H_2O}$. As shown in Fig.\,\ref{fig:control}, one can slow down the voltage ramp to a ramp time of $1\tau_{\rm m}^{\rm fuel}$ to reduce most of the current undershoot induced by the mass-transfer lag. While it is difficult to reduce the undershoot induced by heat-transfer lag by simply slowing down the voltage ramp due to the large $\tau_{\rm h}$. Overall, the characteristic time derived in this study improves the generalizability of control strategies by providing a quantifiable indicator that is applicable to different SOCs.

\section{Conclusions} \label{Sec:Conclu}
The transient characteristics of SOCs are heavily influenced by heat and mass transfer processes within the cell. In this study, we applied non-dimensional analysis to identify parameters and time constants that directly influence SOC transients. We conducted a parametric analysis using 3-D numerical simulation to investigate gaseous and thermal responses of SOCs with varying dimensions, material properties, and operating conditions.  Based on simulation results, we proposed the characteristic times $\tau_{\rm h}$ and $\tau_{\rm m}$ for the first time to represent the overall heat and mass transfer rates within various SOCs under different operating conditions. Gaseous and thermal response times upon rapid electrical variations are at the order of $\tau_{\rm m}$ and $\tau_{\rm h}$, respectively. The effectiveness of $\tau_{\rm h}$ and $\tau_{\rm m}$ was also validated against the transient characteristics of multiple types of SOCs and a PEMFC reported in the literature. In terms of applications, on one hand, $\tau_{\rm h}$ and $\tau_{\rm m}$ can serve as quantifiable indicators for designing an SOC with desired transient characteristics through simple calculations. On the other hand, $\tau_{\rm m}$ can enhance generalizability of existing control strategies. The proposed characteristic times improve theoretical understandings on SOC transients and boost development of SOC for renewable energy storage and management. The methodology and conclusions presented in this study are beneficial for characterizing dynamic processes of other electrochemical cells.

\appendix

\section*{Acknowledgement} 
The authors gratefully acknowledge the partial support from the Hong Kong Polytechnic University Grant P0035016. We would like to acknowledge the use of ChatGPT in improving the writing of this paper. We are grateful to the team behind ChatGPT for their work in developing this tool.

%---------------Insert List of symbols--------------------
%-----------------List of symbols-------------
\mbox{}
%---------------Variables--------------
\nomenclature[A]{$\delta$}{Thickness, [m]}
\nomenclature[A]{$H$}{Height, [m]}
\nomenclature[A]{$L$}{Length, [m]}
\nomenclature[A]{$W$}{Width, [m]}
\nomenclature[A]{$T$}{Temperature, [K]}
\nomenclature[A]{$p$}{Pressure, [Pa]}
\nomenclature[A]{$\Vec{V}$}{Velocity, [m/s]}
%\nomenclature[A]{$V$}{Volume, [m$^3$]}
\nomenclature[A]{$\dot{V}$}{Volumetric flow rate, [m$^3$/s]}
\nomenclature[A]{$V_{\rm in}$}{Inlet velocity, [m/s]}
\nomenclature[A]{$t$}{Time, [s]}
\nomenclature[A]{$Y_i$}{Mass fraction of species $i$}
\nomenclature[A]{$k$}{Thermal conductivity, [W/m.K]}
\nomenclature[A]{$c_p$}{Thermal capacity, [J/kg.K]}
\nomenclature[A]{$\varepsilon$}{Porosity}
\nomenclature[A]{$S$}{Source term}
\nomenclature[A]{$X_i$}{Mole fraction of species $i$}
\nomenclature[A]{$\alpha$}{Thermal diffusivity, [m$^2$/s]}
\nomenclature[A]{$R$}{Ideal gas constant, [J/mol.K]}
\nomenclature[A]{$\rho$}{Density, [kg/m$^3$]}
\nomenclature[A]{$\tau$}{Time constant, [s]}
\nomenclature[A]{$D$}{Diffusivity, [m$^2$/s]}
\nomenclature[A]{$\phi_{\rm ele}$}{Electrical potential, [V]}
\nomenclature[A]{$i$}{Current density, [A/cm$^2$]}
\nomenclature[A]{$m$}{Mass, [kg]}
\nomenclature[A]{$n$}{Mole, [mol]}
\nomenclature[A]{$\dot{\mathcal{H}}$}{Enthalpy flow rate, [W]}

%-------------Subscripts---------------
\nomenclature[B]{f}{Fluid}
\nomenclature[B]{s}{Solid}
\nomenclature[B]{ch}{Fluid channel}
\nomenclature[B]{th}{Thermal}
\nomenclature[B]{m}{Mass}
\nomenclature[B]{0}{Reference value}
\nomenclature[B]{eff}{Effective}
\nomenclature[B]{h}{heat}
\nomenclature[B]{Int}{Interconnect}

%-------------Abbreviation-----------------
\nomenclature[C]{SOC}{Solid oxide cell}
\nomenclature[C]{SOEC}{Solid oxide electrolysis cell}
\nomenclature[C]{SOFC}{Solid oxide fuel cell}
\nomenclature[C]{AFL}{Anode functional layer}
\nomenclature[C]{CFL}{Cathode functional layer}
\nomenclature[C]{ADL}{Anode diffusion layer}
\nomenclature[C]{CDL}{Cathode diffusion layer}
%\nomenclature[C]{TPB}{Triple phase boundary}

\printnomenclature
%-----------------List of symbols-------------

\bibliographystyle{elsarticle-num} 

\biboptions{sort&compress} 

\bibliography{Refs}

\end{document}